\documentclass{article}

\usepackage[a4paper]{geometry}
\usepackage{amssymb}
\usepackage{amsmath}
\usepackage{mathtools}
\usepackage[sectionbib]{natbib}

\usepackage{eurosym}
\usepackage{amsthm}
\usepackage{caption}
\usepackage{comment}
\usepackage[utf8]{inputenc} % allow utf-8 input
\usepackage[T1]{fontenc}    % use 8-bit T1 fonts
\usepackage{hyperref}       % hyperlinks
\usepackage{url}            % simple URL typesetting
\usepackage{booktabs}       % professional-quality tables
\usepackage{amsfonts}       % blackboard math symbols
\usepackage{nicefrac}       % compact symbols for 1/2, etc.
\usepackage{microtype}      % microtypography
\usepackage{cleveref}
\usepackage[font=small,skip=10pt,labelfont=bf,tableposition=top]{caption}
\usepackage[font=small,skip=0.3pt]{subcaption}
\usepackage[section]{placeins}
\usepackage{array}

\newcount\Comments  % 0 suppresses notes to selves in text
\usepackage{color}
\definecolor{darkgreen}{rgb}{0,0.5,0}
\definecolor{purple}{rgb}{1,0,1}
\newcommand{\kibitz}[2]{\ifnum\Comments=1\textcolor{#1}{#2}\fi}
\usepackage{booktabs}

\usepackage{multirow}
\makeatletter
\newcommand\thefontsize{The current font size is: \f@size pt}
\makeatother

\usepackage{setspace}

\DeclareCaptionLabelFormat{andtable}{#1~#2  \&  \tablename~\thetable}

 \title{SpotV2Net:\\ Multivariate Intraday Spot Volatility Forecasting \\ via Vol-of-Vol-Informed Graph Attention Networks}

\author{Alessio Brini$^{a}$\footnote{Corresponding author.}  \qquad Giacomo Toscano$^{b}$}

\date{}

\begin{document}

\maketitle
 \vspace{3cm}
\thispagestyle{empty} \medskip
 \date{ \medskip \medskip $^a$\textit{Duke University Pratt School of Engineering, 305 Teer Engineering Building Box 90271} \\  \centering \textit{Durham, NC 27708 (USA).
 E-mail: alessio.brini@duke.edu}\medskip \\    \indent  
 $^b$\textit{Department of Economics and Management, University of Florence, Via delle Pandette 32}  \\   \centering 
 \textit{Firenze, 50127 (Italy). E-mail: giacomo.toscano@unifi.it}  \medskip \\ 
 }
\bigskip
\indent  \medskip


\begin{abstract}
This paper introduces SpotV2Net, a multivariate intraday spot volatility forecasting model based on a Graph Attention Network architecture. SpotV2Net represents assets as nodes within a graph and includes non-parametric high-frequency Fourier estimates of the spot volatility and co-volatility as node features. Further, it incorporates Fourier estimates of the spot volatility of volatility and co-volatility of volatility as features for node edges, to capture spillover effects. We test the forecasting accuracy of SpotV2Net in an extensive empirical exercise, conducted with the components of the Dow Jones Industrial Average index. The results we obtain suggest that SpotV2Net yields statistically significant gains in forecasting accuracy, for both single-step and multi-step forecasts, compared to a panel heterogenous auto-regressive model and alternative machine-learning models. To interpret the forecasts produced by SpotV2Net, we employ GNNExplainer \citep{ying2019gnnexplainer}, a model-agnostic interpretability tool, and thereby uncover subgraphs that are critical to a node's predictions.
\end{abstract}
 
\newpage

\section{Introduction}\label{Sec:intro}

Volatility forecasting has long been at the core of the literature on financial econometrics,  as scholars and practitioners have been continually seeking accurate and robust asset volatility forecasts for applications in different areas, including option pricing, portfolio optimization and financial risk management (see, respectively, \cite{christoffersen2000relevant},  \cite{bandi2008realized} and \cite{becker2015selecting}, among many others).

Even though the focus of academic research has mainly been on forecasting one-day-ahead volatilities (see, for instance, \cite{brailsford1996evaluation,poon2003forecasting,andersen2006volatility,satchell2011forecasting}), in recent years, the fast development of the high-frequency trading industry has emphasized the need for reliable intraday volatility forecasts, see \cite{engle2012forecasting,rossi2015long,zhang2023volatility}. The availability of accurate intraday volatility forecasts can be essential for the decision-making process of market operators in different circumstances.

Firstly, short-term volatility forecasts may be exploited to perform timely risk management tasks within the daily horizon (see, e.g., \cite{madsam2019intradayrisk, Rice2020ForVaR}). Indeed, intraday volatility prediction can anticipate sudden market shifts, especially after significant events, and provide valuable information for traders and portfolio managers, who rely on real-time risk assessments to adjust positions or hedge strategies promptly. Secondly, with the growth of algorithmic and high-frequency trading, modeling and forecasting intraday volatility dynamics has become foundational to guide profit-making strategies that hinge on short-term price movements (see, e.g., \cite{Liu2018tradhfvol,BEHRENDT2018355,goldstein2023hftradstrat,mariotti2023fromzero}). For instance, anticipating short-term volatility changes could allow high-frequency traders to improve optimal execution, exploit sudden shifts in market sentiment and optimally plan market entrance/exit. At the same time, financial institutions may exploit accurate short-term volatility predictions to timely adjust trading margins. Similarly, market makers could use intraday volatility forecasts to optimally plan their operations in order to control bid-ask spreads and guarantee adequate liquidity.

Beyond the trading setting, anticipating intraday volatility changes may also be instrumental for the early detection of anomalies at the microstructural level (see, for example, \cite{Xue2012freqclust,Ligot2021intradaysmile}). Moreover, regulators could use short-term volatility forecasts to anticipate the outset of instability conditions \citep{ALLAJ20231777}.

In this paper, we introduce a novel approach to multivariate intraday volatility forecasting based on Graph Neural Networks (GNNs). GNNs represent a specific instance of neural network architecture that can act on data structured as graphs and have been successfully employed to model several complex multidimensional dynamic problems in different fields, such as traffic flows  \citep{diao2019dynamic,li2021spatial,jiang2022graph}, recommendation systems \citep{ying2018graph,huang2021mixgcf,xia2022multi}, social networks \citep{fan2019graph,min2021stgsn}, supply chains  \citep{gopal2021discovering},   international trade \citep{panford2020bilateral,monken2021graph}, interbank markets \citep{liu2023preventing}, enterprise bankruptcy \citep{zhao2022combining} and tax evasion \citep{shi2023edge}.
The dynamics of asset volatilities in a high-dimensional intraday setting may be seen as a complex system, %of connected entities, 
based on the empirical evidence that suggests the presence of intraday co-movements and spillovers effects across different sectors and markets, see, e.g., \cite{Golos2015intradayspill,jaw2015bidirectional, Nishi2018spillindex,FASSAS2019333,KATSIAMPA201935,NAEEM2023106677}. Hence, GNNs can represent an effective modeling framework to capture the changing patterns in the intraday volatility of assets by leveraging data from various correlated securities.

In a multidimensional setting, we assume that each asset represents a node of a graph. The features (covariates) associated with each node comprise the instantaneous (or spot) volatility of the asset and the instantaneous co-volatilities with every other asset in the graph, reconstructed using the non-parametric Fourier methodology by \cite{malliavin2002fourier,malliavin2009fourier}. Furthermore, to capture spillover effects, the edge features encapsulate information about the dynamic dependence between node pairs. Specifically, for any given edge, its features consist of Fourier estimates of the volatility of volatility of each of the assets connected by the edge and the co-volatility of their individual volatilities.  In this regard, we note that different studies have shown that encompassing realized or implied volatility of volatility measures improves a model's performance in risk management tasks (see, e.g., \cite{Huang_Schlag_Shaliastovich_Thimme_2019,chen2021asset,LI2022energy} and forecasting applications (see, e.g., \cite{CATANIA2020volvol,CAMPISI2023volvol,DING2023joe}). Node and edge features represent the inputs of the forecasting model. To account for (cross) serial dependence, each input includes not only current values but also several lagged values. This is achieved by leveraging the capability of deep learning architectures to handle a large number of inputs \citep{jaegle2021perceiver,menghani2023efficient}. In the specific case of GNNs, inputs can be placed both in the nodes and the edges \citep{wu2020comprehensive,zhou2020graph}. We refer to the resulting architecture that we obtain as Spot Volatility and Volatility of volatility Network, abbreviated to SpotV2Net.

The GNN architecture that we employ in this paper relies on the attention mechanism and thus falls within the Graph Attention Network (GAT) framework proposed by \citep{velivckovic2017graph}. The attention mechanism allows assigning different attention scores to different nodes in a graph, thereby enabling the forecasting model to attribute more weight to the information embedded into a specific subset of neighbor nodes. This adaptability can be crucial for the efficiency of empirical applications in which nodes are not likely to contribute equally to the forecast of the volatility of a given asset in a graph. Moreover, the GAT framework can inherently handle edge features because the attention mechanism includes also the latter in the calculation of attention scores. On the contrary, other types of GNN instances, such as Graph Convolutional Networks (GCNs) \citep{kipf2016semi}, Spectral-based GNNs \citep{defferrard2016convolutional} and Spatial-based GNNs \citep{hamilton2017inductive}, operate with a fixed, predetermined weight for neighbor nodes and do not naturally handle edge features.

We test the SpotV2Net model on the universe of the 30 stocks that compose the Dow Jones Industrial Average (DJIA) index at the time of writing. Using 1-second asset prices, we estimate model inputs, namely univariate and multivariate spot volatility and volatility of volatility time series, on the 30-minute grid. Then, we evaluate the performance of SpotV2Net in performing single-step and multi-step spot volatility forecasts and compare its accuracy with that of alternative models. These include a Heterogeneous Auto-Regressive (HAR) model for the spot volatility\footnote{We adapt the HAR model by \cite{corsi2009simple}, which is built to forecast daily integrated volatilities, to forecast intraday spot volatilities. The resulting model, which maintains a heterogeneous auto-regressive structure, is termed HAR-Spot and is illustrated in Appendix \ref{sec:appendix_alternativemodels}.} and two machine learning (ML) models that do not handle graph structures, namely the Extreme Gradient Boosting (XGB) model and the Long Short-Term Memory (LSTM) model. We find that the capability of SpotV2Net to learn from graph-like structured data and, in particular, to capture spillover effects via edge features, yields statistically significant gains in forecasting accuracy with respect to the alternative models considered. Such gains are obtained both with single-step and multi-step forecasts. In this regard, we note that SpotV2Net is capable of producing multi-step forecasts without resorting to a recursive approach.

Besides, given the complexity of the GAT architecture employed, we also address the issue of the interpretability of the predictions produced by the SpotV2Net model. Specifically, we employ the GNNExplainer framework \citep{ying2019gnnexplainer} to identify relevant nodes and edges that influence the model's spot volatility predictions for a specific asset.

The paper is organized as follows. Sec. \ref{Sec:literature} resumes the relevant literature on multivariate volatility forecasting, while Sec. \ref{Sec:notation} introduces the notation and illustrates the assumptions behind the estimation of univariate and multivariate spot volatility and volatility of volatility time series. Sec. \ref{Sec:motivation} motivates the use of a GAT framework with co-volatilities of volatilities as edge features for forecasting intraday multivariate spot volatilities.
Sec. \ref{Sec:model} then formally introduces the SpotV2Net model, while Sec. \ref{Sec:estimators} recalls the definition of the Fourier spot estimators employed to reconstruct the inputs of SpotV2Net. Further, Sec. \ref{Sec:results} illustrates the empirical application of SpotV2Net to the universe of DJIA stocks and Sec. \ref{Sec:conclusion} concludes. Finally, App. \ref{sec:appendix_alternativemodels} briefly illustrates the modeling alternatives to SpotV2Net which are implemented for comparison in the empirical application of Sec. \ref{Sec:results}, while App. \ref{Sec:appendix_hyperparameters} provides information on the setting of hyperparameters.

\section{Related literature}\label{Sec:literature}

The complexity of financial markets explains the growing interest of the literature on multivariate models that can capture volatility comovements among different securities. In this regard, see, for instance, \cite{bauwens2006multivariate}, who provide a comprehensive survey on the multivariate extension of GARCH-like models - such as the diagonal VEC model \citep{bollerslev1988capital}, the BEKK model \citep{engle1995multivariate}, the VAR-GARCH model \citep{ling2003asymptotic} and the Wishart Autoregression (WAR) \citep{gourieroux2009wishart}- and \cite{wilms2021multivariate}, who employ a multivariate version of the HAR model \citep{corsi2009simple} to capture volatility spillover effects among stock market indices. 

\cite{herskovic2016common,caldeira2017combining,bollerslev2019high,herskovic2020firm} show the importance of these models for applications in asset pricing, portfolio optimization and risk management. However, these models present a number of drawbacks. For example, the diagonal VEC, the BEKK and the HAR face significant computational challenges when scaling to a large number of assets. For instance, \cite{callot2017modeling} explore the curse of dimensionality in multivariate volatility models, noting that these models may lead to poor out-of-sample forecasts and high computational burdens, making the development of a more scalable solution a valuable approach in this area of study. Additionally, as pointed out by \cite{zhang2022graph,zhang2023graph}, VAR-GARCH and WAR struggle to capture long-term dependencies in volatility time series.

Within a multivariate framework, forecasting an intraday spot measure of asset volatility is a critical area of exploration that has received little attention in the literature, to the best of our knowledge. Few works attempt to model intraday volatility through ARCH  \citep{taylor1997incremental} and GARCH \citep{engle2012forecasting} variants. However, their modeling focus is on integrated measures of volatilities, namely on realized volatilities \citep{andersen2001distribution}, rather than on spot estimates. \textcolor{black}{Moreover, as pointed out by \cite{christensen2023machine}, traditional linear models may break down when the explanatory variables are strongly correlated, exhibit low signal-to-noise
ratios, or if the underlying structure is nonlinear.}

\textcolor{black}{On this behalf, ML techniques, which have recently made an impact on the field of finance, represent an alternative approach to effectively handling these issues.} The capability of ML models, especially deep neural networks, to handle high-dimensional data and approximate complex patterns has impacted key areas of finance, such as portfolio optimization \citep{heaton2017deep,zhang2020deep,ma2021portfolio,lin2023portfolio}, asset pricing \citep{gu2020empirical,wu2021equity2vec,chen2023deep} and volatility forecasting \citep{xiong2015deep,liu2019novel,bucci2020realized,christensen2023machine,zhu2023forecasting}. For the specific problem of volatility forecasting,  it is worth underlying that the findings by \cite{christensen2023machine} suggest that several ML
algorithms, despite being implemented with minimal hyperparameter tuning, yield improved forecasting performance compared to models in the HAR lineage, especially at longer horizons. The authors
attribute such forecast accuracy gains to higher persistence in the ML models - which helps to approximate
the long memory of realized variance - and the capability of ML models to encompass the contribution from additional predictors, besides past volatility lags.

GNNs are part of the ML-based approaches that can play a significant role in enhancing financial applications, see, e.g., the review by \cite{wang2021review} and \cite{zhang2022research}. Recent studies (see \cite{chen2022multivariate,reisenhofer2022harnet,zhang2022graph,djanga2023cryptocurrency}) employ GNN-based models to forecast integrated measures of asset volatility, illustrating the potential of graph-based methods in capturing complex inter-asset relationships, including non-linear volatility spillover effects \citep{zhang2023graph}. Similarly, \cite{wu2020connecting} and \cite{cheng2022financial} extend the scope of application for GNNs by employing the latter for forecasting financial time series. Building on these foundations, our research aims to extend the use of GNNs by including the attention mechanism typical of GATs. GAT architectures have already been employed to predict stock market movements \citep{kim2019hats,cheng2022financial} and provide stock recommendations \citep{ying2020time}. However, few works have leveraged the flexibility of GATs for modeling spillover effects in financial markets. For instance, \cite{cheng2021modeling} model the momentum effect between firms using an attribute-sensitive architecture similar to our approach, but they focus on solving a binary classification problem to predict the directional movements of stock prices, rather than forecasting asset volatilities.

\section{Notation and assumptions}\label{Sec:notation}

We consider a collection of $N$ assets. For $i=1,...,N$, we let $p_i(t)$ and $V_i(t)$ denote, respectively, the log-price and the spot volatility of $i$-th asset at time $t \in [0,T]$. We assume that the dynamics of the pair $(p_i(t), V_i(t))$ satisfy 
\begin{equation*}
\begin{cases}
 & dp_i(t)=   \mu_{i,1}(t) dt +   \sqrt V_i(t) dW_{i,1}(t)\\
&   dV_i(t)=   \mu_{i,2}(t) dt +   \sqrt{\widetilde V}_i(t) dW_{i,2}(t)
 \end{cases} \, ,
 \end{equation*}
where, for any $i \in \{1,\dots, N\}$ and for any $k\in \{1,2\}$, $W_{i,k}$ is a Brownian motion on the filtered probability space $(\Omega, P, \mathcal{F}, (\mathcal{F}_t)_{t\in[0,T]})$, while the drift $\mu_{i,k}$ and the spot volatility of volatility $\widetilde V_i$ are continuous, bounded and adapted stochastic processes defined on $(\Omega, P, \mathcal{F}, (\mathcal{F}_t)_{t\in[0,T]})$. The pairwise correlations between the Brownian motions $W_{1,1},...,W_{N,1}, W_{1,2},...,W_{N,2}$ are allowed to be time-dependent and random.

Let $\langle p_i, p_j  \rangle_t$ denote the quadratic covariation between $p_i$ and $p_j$ on $[0,t]$.
We define the spot co-volatility process as  
$$C_{ij}(t):=\frac{d\langle p_i, p_j \rangle_t}{dt}$$
and note that $C_{ii}(t)=V_i(t)$. Similarly, we define the spot co-volatility of volatility process as
$$ \widetilde C_{ij}(t):=\frac{d\langle V_i, V_j \rangle_t}{dt},$$
with $\widetilde C_{ii}(t)=\widetilde V_i(t)$.

The SpotV2Net model employs estimates of the discrete trajectories
$$  \{ C_{ij,t} \}_{t \in \mathcal{T}}, \quad \{ V_{i,t} \}_{t \in \mathcal{T}}, \quad\{ \widetilde C_{ij,t} \}_{t \in \mathcal{T}}, \quad\{ \widetilde V_{i,t} \}_{t \in \mathcal{T}},\quad   (i,j) \in \{1,...,N\}\times\{1,...,N\}, \, i \neq j,  $$
where $\mathcal{T} =\{0=\tau_0 < \tau_1 < \ldots < T=\tau_B \}$, $B \in \mathbb{N}$. For ease of notation, we use, e.g., $V_{i,b}$ to indicate $V_{i, \tau_b}$, $b=0,1, \dots, B$. Estimates are obtained using the Fourier methodology, which is detailed in Sec. \ref{Sec:estimators}. We denote the estimate of, e.g., $V_{i,b}$, by $\widehat V_{i,b}$.

\section{Motivation}\label{Sec:motivation}

In the last two decades, different studies have highlighted the importance of including time-varying volatility of volatility effects in financial models to accurately capture price and volatility stylized facts and improve forecasting performance. For instance, in their seminal work, \cite{BTZ} found that accounting for time-varying volatility of volatility effects is of primary importance for modeling and forecasting the temporal variation of expected returns. 
Furthermore, the findings by \cite{cmpp} suggest that allowing for time-varying volatility of volatility substantially improves both the fit and the predictive performance of traditional autoregressive models like the ARFIMA \citep{granger1980introduction} and the HAR models.
More recently, other works have demonstrated the importance of the volatility of volatility in forecasting financial variables.   \cite{CATANIA2020volvol}  propose a score-driven model for the realized
volatility and price dynamics with time-varying volatility of volatility effects and show that the latter contributes to improving out-of-sample performance, compared to a benchmark HAR-GARCH model. Additionally, \cite{DING2023joe} introduces a GARCH-type model that allows for conditional heteroskedasticity in the volatility of asset returns and shows that this model can capture returns stylized facts and yield more accurate volatility forecasts, compared to commonly-used models in the GARCH family. Moreover,  beyond traditional autoregressive models, \cite{CAMPISI2023volvol} found, by implementing several ML methods, that the implied volatility of volatility, as proxied by the CBOE VVIX index\footnote{\href{https://www.cboe.com/us/indices/dashboard/vvix}{CBOE VVIX website}.},
is a predictor, along with other market indices, of the future direction of the stock market.

 \bigskip
 
 The SpotV2Net model employs the information embedded in spot co-volatility of volatility paths to obtain accurate intraday spot volatility forecasts in a multivariate setting. Specifically, the model rests on the idea that instantaneous spillover effects between asset volatilities are naturally captured by the dynamics of the time-varying spot co-volatilities of volatility pairs. 

 Intuitively, SpotV2Net views asset spot volatilities at a given point in time as different nodes of a graph that are connected by edges embedding information about the contingent \textit{strength} of the relationship between nodes' pairs.  Such information is naturally represented by 
 the instantaneous dependence between two nodes (i.e., volatilities), as measured by the spot co-volatility between the nodes (i.e., the spot co-volatility of volatility). Given this structure, a time-varying co-volatility of volatility can be naturally seen as an indicator of the magnitude of instantaneous spillover effects.

 We can illustrate the role of the co-volatility of volatility with an empirical example. Consider the two
consequential financial events that occurred in March 2023, namely the collapse of Silicon Valley Bank (on March 10) and Signature Bank (on March 12). Given the pivotal role of banks in the financial system, these bank failures immediately initiated a volatility burst that spread from the US to international markets. However, not all stocks' volatilities were affected by spillover effects of the same magnitude.

\begin{figure}
    \centering
    \includegraphics[scale=0.85]{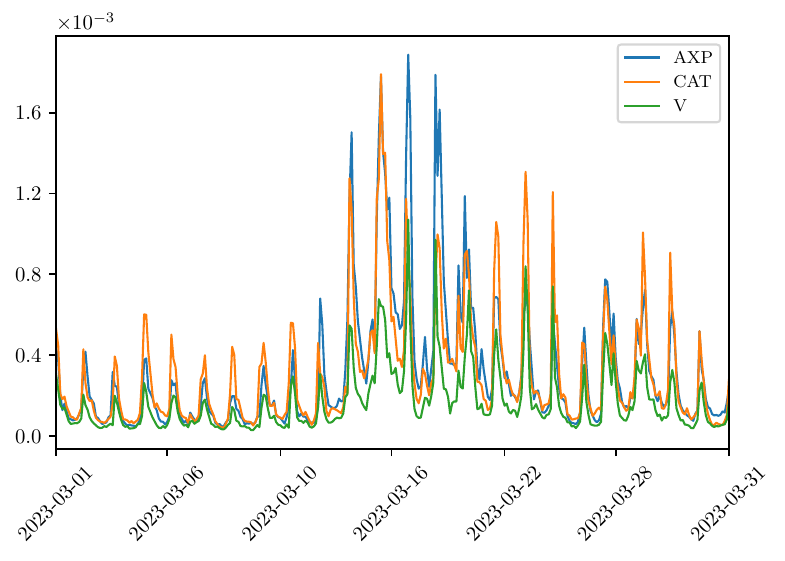} 
    \caption{\textcolor{black}{Fourier spot volatility estimates $\{ V_{i,t} \}_{t \in \mathcal{T}}$ for $i \in \{AXP, CAT, V\}$, on the 30-minute grid $\mathcal{T} = \{\tau_j=jT/13 : \,  j=0,1,...,13\}$, where $T$ is a 6.5-hour trading day. Estimates were obtained using 1-second prices.
Details about the Fourier estimator of the spot volatility are provided in Sec. \ref{Sec:estimators}.}}
    \label{fig:vols_march}
\end{figure}

 For example, consider the following three US companies, that are part of the Dow Jones Industrial Average index: American Express (AXP) and Visa (V), two companies that provide financial services, 
and Caterpillar (CAT), a manufacturing company. As illustrated in Fig. \ref{fig:vols_march}, all three companies experienced an increase in their spot volatility level right after the failures of Silicon Valley Bank and Signature Bank, showing the largest volatility peaks right after March 13. The largest volatility peaks were then followed by sequences of local peaks of lower magnitude and a progressive reversion to the volatility level observed at the beginning of the month.

\begin{figure}
    \centering
    \includegraphics[scale=0.85]{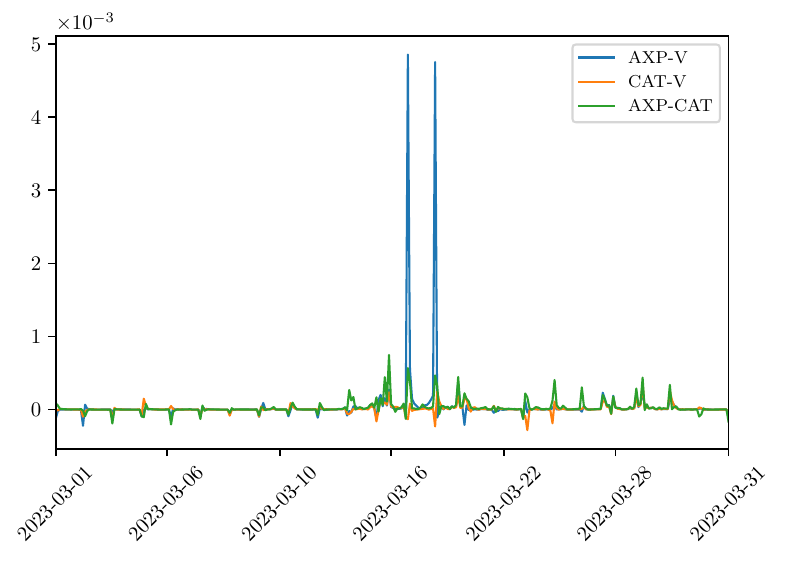} 
\caption{\textcolor{black}{Fourier spot co-volatility of volatility estimates $\{ \widetilde{C}_{ij,t} \}_{t \in \mathcal{T}}$ for $i,j \in \{AXP, CAT, V\}, \, i \neq j$, on the 30-minute grid $\mathcal{T} = \{\tau_j=jT/13 : \,  j=0,1,...,13\}$, where $T$ is a 6.5-hour trading day. Estimates were obtained using 1-second prices.
Details about the Fourier estimator of the spot volatility are provided in Sec. \ref{Sec:estimators}.}}
    \label{fig:covolvols_march}
\end{figure}

 However, by observing Fig.  \ref{fig:covolvols_march}, one immediately notices that the pairwise co-volatilities of volatility behave very differently. In particular, the spot co-volatility of volatility between
AXP and V shows two very large peaks near March 16 and March 17, while the co-volatility of volatility between each of these two stocks and CAT maintains substantially stable and lower values during March 2023. Such bursts in the spot co-volatility of volatility may indicate the presence of more pronounced instantaneous spillover effects between AXP  and V, compared to the pairs AXP-CAT and V-CAT. This is not surprising, since AXP and V operate in the same sector, namely, financial services, which the collapse of Silicon Valley Bank and Signature Bank directly impacted. Instead, CAT is a well-diversified manufacturing company whose spot volatility may be less correlated with the volatility of companies operating in the banking and insurance industry. SpotV2Net can recognize such effects using the attention mechanism inherent in the GAT architecture. When predicting the future value of a specific node, the attention mechanism allows assigning different weights to neighbor nodes, based not only on node features but also on features embedded in the connecting edges. Hence, the attention mechanism enables SpotV2Net to naturally capture the time-varying spillover effects between volatility pairs by including edge features. The next Section describes the attention mechanism in detail.

% \FloatBarrier

% \FloatBarrier

\section{The SpotV2Net Model}\label{Sec:model}
 
 GNNs are a family of deep learning models \citep{goodfellow2016deep} that can learn from data structured as graphs. They differ from other widely-used neural network architectures such as Convolutional Neural Networks (CNNs) \citep{lecun1995convolutional} and Recurrent Neural Networks (RNNs) \citep{rumelhart1985learning,hochreiter1997long}, which are designed to model for grid and sequence data, respectively.

The general idea of GNNs rests on the principle of iterative message passing, where graph nodes aggregate information from themselves and their connected nodes, referred to as neighbors. This aggregation can be repeated by stacking multiple layers, where each layer generates its own representation based on the output of the previous layer. These representations are then passed to a linear aggregation layer to perform node regression, i.e., obtaining output forecasts for each node. This end-to-end learning process allows training the model to operate directly on graph-structured data.  

The mathematical operation that produces the aggregation defines the type of GNN algorithm. \cite{scarselli2008graph} present the first example of GNN modeling characterized by an iterative updating procedure for the node representation.  
Other models introduce improvements in computational efficiency, ease of training and the ability to handle complex graph structures. For instance, Graph Convolutional networks (GCN) allow the computation to be performed in parallel and follow a static layered structure similar to a feedforward neural network \citep{kipf2016semi}. Furthermore, Graph Isomorphism Network (GIN) \citep{xu2018powerful} includes fixed trainable weights that determine the different impacts of neighboring nodes when performing the prediction of a specific node.

The GAT \citep{velivckovic2017graph}, on which SpotV2Net is based, introduces an attention mechanism that sets it apart from other models. While GCNs propagate and aggregate information uniformly from neighbors, and GIN ensures that the network can effectively discern the relative significance of a node's inherent features as opposed to the aggregated features from neighboring nodes, GAT empowers nodes to assign different weights to their neighbors based on their features and the features of the connecting edges. This dynamic mechanism builds on transformer-based natural language processing models \citep{vaswani2017attention}.

Let us now consider the mechanism behind a GAT. Let $\mathcal{G} =\{\mathcal{V} , \mathcal{E} \}$ denote a graph structure,  where $\mathcal{V} =\left\{\nu_{1}, \ldots, \nu_{N} \right\}$ is the set of the $N$ nodes of the structure and $\mathcal{E}$ is the set of edges, where $\varepsilon_{i j}=\left(\nu_{i}, \nu_{j}\right) \in \mathcal{E}$ denotes the edge connecting node $\nu_{i}$ and node $\nu_{j}$. Besides, 
let ${A}$ denote the $N\times N$ binary adjacency matrix of the structure, where ${A}[i, j]$ is equal to $1$ (respectively, $0$) if   $\nu_{i}$ and $\nu_{j}$ are connected (respectively, are not connected).
The graph $\mathcal{G}$ is assumed to be fully connected, that is, all the entries of $A$ are assumed to be equal to $1$.

A GAT involves a sequence of layers, where the previous layer's input represents the next layer's output. The final layer is referred to as the prediction (output) layer. Intermediate layers are referred to as hidden layers.
When implementing SpotV2Net, we optimize the number of hidden layers as a hyperparameter, determining that two hidden layers are optimal. 

The input to the first GAT layer is the set of vectors representing node features, namely $x = \{ x_1, x_2, \ldots, x_N \}$, $x_i \in \mathbb{R}^M$, where  $M$ indicates the number of features in each node. The GAT layer outputs a different representation of the initial set of node features, with different cardinality $M'$,  
denoted by $x' = \{ x'_1, x'_2, \ldots, x'_N \} $, $x'_i \in \mathbb{R}^{M'}$. We treat $M^\prime$ as a hyperparameter and optimize its value through the training phase. 

Specifically, the operation performed by the GAT layer consists of the following steps. Firstly, the attention coefficient that indicates the importance of node $j$'s features to node $i$, denoted by $e_{ij}$, is obtained as
\begin{equation*}
e_{ij} = a(W x_i, W x_j),
\end{equation*}
where $a$ is a $\mathbb{R}$-valued function defined on $\mathbb{R}^{M'} \times \mathbb{R}^{M'}$ and $ W \in \mathbb{R}^{M' \times M}$ is a weight matrix.
Then, attention coefficients are normalized to make the values comparable across nodes. The normalization is performed using the $softmax$ function, that is, 
\begin{equation*}
\alpha_{ij} = \text{softmax}_j(e_{ij}) = \frac{\exp(e_{ij})}{\sum_{k=1}^N \exp(e_{ik})}.
\end{equation*}

In our study, we follow the implementation by \cite{velivckovic2017graph}, so that $a$ is a single-layer feedforward neural network with LeakyReLU activation function \citep{maas2013rectifier} and weight vector $q \in \mathbb{R}^{2M'}$. The LeakyReLU function is defined as 
\begin{equation*}
\mathcal{L}(x) =
\begin{cases}
x & \text{if } x > 0  \\
c x & \text{if } x \leq 0 
\end{cases}, 
\end{equation*}
where $c$ is a positive constant. The resulting attention coefficient reads as

\begin{equation*}
\alpha_{ij} = \frac{\exp ( \mathcal{L} ( q^T [W x_i \parallel W x_j] ))}{\sum_{k=1}^N \exp ( \mathcal{L} ( q^T [W x_i \parallel W x_k] ))},
\end{equation*}
where   $\cdot^T$ represents transposition and $\parallel$ indicates the concatenation operation.

Once the attention parameters are computed, the hidden representation for each node in the graph is expressed as
\begin{equation*}
x'_i = \sigma \left( \sum_{j=1}^N \alpha_{ij} W x_j \right),
\end{equation*}
where $\sigma$ represents a nonlinear activation function chosen as a hyperparameter of the model.

To improve the expressiveness \footnote{Expressiveness (or capacity) refers to the model's ability to capture complex patterns and relationships in the data. For more details, see Chapter 5 of \cite{goodfellow2016deep}.} and accuracy of the training process \citep{vaswani2017attention}, when implementing SpotV2Net, we follow a multi-head attention mechanism, which consists of a number $K$ of independent attention mechanisms. Multiple heads perform the same operations on the same input but are initialized with different weights. 
 This approach allows each head to learn distinct patterns from the inputs and capture various aspects of the data.  
The resulting hidden node representation is obtained via concatenation and reads as

\begin{equation*}
x'_i = \big\|_{k=1}^{K} \,\, \sigma \left( \sum_{j =1}^N \alpha^k_{ij} W^k x_j \right),
\end{equation*}
where the $\alpha^k_{ij}$'s are normalized attention coefficients computed by the $k$-th attention mechanism and $W^k$ is the corresponding weight matrix. 
To avoid an excessive increase in dimensionality, which could lead to overfitting \citep{velivckovic2017graph}, concatenation is not performed if the hidden layer precedes the final prediction layer. Instead, in such case, an average is computed, that is,
\begin{equation*}
x'_i = \sigma \left( \frac{1}{K} \sum_{k=1}^{K} \sum_{j =1}^N \alpha^k_{ij} W^k x_j \right).
\end{equation*}

In the last layer, an affine transformation is applied to obtain the prediction. Specifically, the final layer yields the prediction for node $i$, denoted by $\hat{y}_i$, which reads as
  \begin{equation*}\label{eq:pred}
  \hat{y}_i = O x'_i + u,
  \end{equation*}
where $O \in \mathbb{R}^{N \times M'}$ and $u \in \mathbb{R}^N$ is a vector. 

\bigskip 

In the specific case of SpotV2Net, at a given time $\tau_b$, we feed the GAT the node feature vectors
\begin{equation}\label{eq:x}
x_i = \left[ \left\{\widehat V_{i,b-l} \right\}_{l=0,\ldots,L} \, , \, \left\{ \left\{ \widehat C_{ij,b-l}\right\}_{j=1,\ldots,N; j \neq i}\right\}_{l=0,1,\ldots,L} \right],  
\end{equation}
that is, $x_i$ includes contemporaneous values and all the lags up to $L$ of the estimated volatility of the $i$-th asset and the estimated co-volatility between the $i$-th asset and all the other assets. Note that the number of lags $L$ is not predetermined a priori but is treated as a hyperparameter, subject to tuning (see the hyperparameter `Number of Lags' in App. \ref{Sec:appendix_hyperparameters}). This allows for flexibility in dynamically capturing the long memory property of the spot volatility.

\begin{figure}
    \centering
    \includegraphics[width=0.7\textwidth]{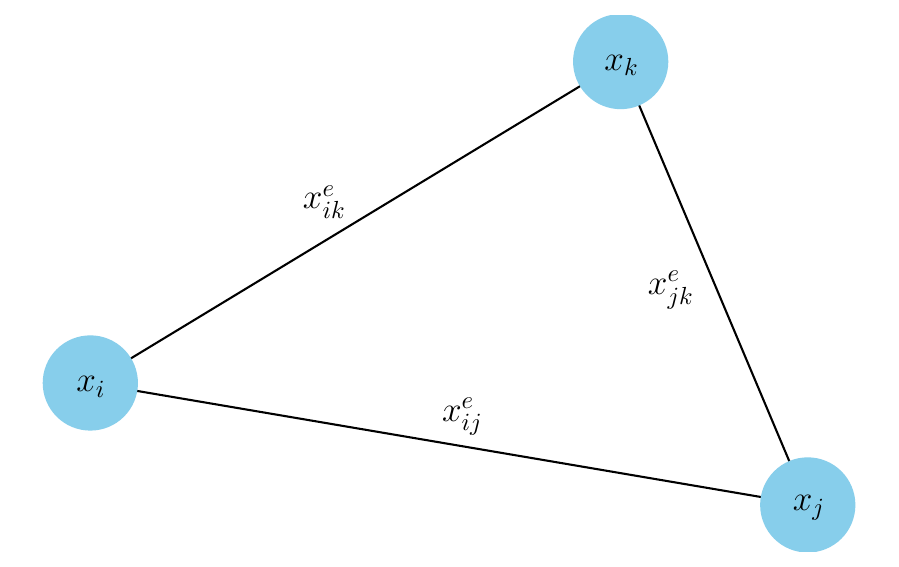} 
    \caption{Three-dimensional example of the fully connected graph topology of SpotV2Net. The three nodes are denoted by $i$,$j$ and $k$. \textcolor{black}{The vectors $x_{*}$,  where $*=i,j,k$, and $x^e_{**}$, where $**=ij,ik,jk$, are defined in Eqns. (\ref{eq:x}) and (\ref{eq:xe}), respectively.}}
    \label{fig:graph}
\end{figure}

Our application of the GAT layer also allows for edge features, i.e., information regarding the dependence between two nodes in a graph. Specifically, we have a set of edge features $x^{e} = \{ x_{ij}^{e} \mid i < j \}, \quad x_{ij}^{e} \in \mathbb{R}^E$. We set
\begin{equation}\label{eq:xe}
x_{ij}^{e} = \left[ \left\{\widehat{\widetilde V}_{i,b-l} \right\}_{l=0,\ldots,L} \, , \, \left\{\widehat{\widetilde V}_{j,b-l} \right\}_{l=0,\ldots,L} \, , \, \left\{ \widehat{\widetilde C}_{ij,b-l}\right\}_{l=0,1,\ldots,L} \right],    \end{equation}
that is, $x_{ij}^e$ includes contemporaneous values and all the lags up to $L$ of the estimated volatility of volatility of the $i$-th asset and the $j$-th asset, along with the estimated co-volatility of volatility between the $i$-th asset and the $j$-th asset. Fig. \ref{fig:graph} illustrates the general structure of SpotV2Net's input graph in the three-dimensional case\footnote{This simplified, low-dimensional representation is intended to illustrate the process clearly, avoiding the cluttered figure that would result from using higher dimensions.  However, the inherent scalability of the GNN architecture allows effortlessly applying SpotV2Net to a much larger cross-section of assets. Accordingly, in the empirical application presented in Sec. \ref{Sec:results}, we employ SpotV2Net to model the spot volatility of all of the 30 constituents of the DJIA index.}.

To include edge features in a GAT, it is sufficient to modify the attention mechanism so that the function $a : \mathbb{R}^{M'} \times \mathbb{R}^{M'} \rightarrow \mathbb{R}$ becomes $a' : \mathbb{R}^{M'} \times \mathbb{R}^{M'} \times \mathbb{R}^{E'} \rightarrow \mathbb{R}$. The resulting normalized attention coefficient then reads as
\begin{equation*}
\alpha'_{ij} = \frac{\exp \left( \mathcal{L} \left( q'^T [W x_i \parallel W x_j \parallel U  {x_{ij}^{e}}] \right) \right)}{ \sum_{k=1}^N {\exp \left( \mathcal{L} \left( q'^T [W x_i \parallel W x_k \parallel U  {x_{ik}^{e}}] \right) \right)}},
\end{equation*}
where $U \in \mathbb{R}^{E' \times E}$ is the weight matrix for the linear transformation of the edge features and $E'$ is the dimension of the transformed edge features.  

\FloatBarrier

\section{Fourier estimators}\label{Sec:estimators}

\textcolor{black}{In this Section, we briefly illustrate the Fourier estimation method, which allows us to efficiently reconstruct the entries of node and edge feature vectors defined, respectively, in (\ref{eq:x}) and (\ref{eq:xe}).}
 The Fourier non-parametric estimation method, originally introduced by \cite{malliavin2002fourier,malliavin2009fourier}, allows reconstructing the latent spot co-volatility paths from high-frequency price observations. The method is particularly apt for estimating spot co-volatilities in the presence of microstructure noise, as it removes the influence of the latter by simply cutting off the highest coefficients of the price increment from the convolution that yields the co-volatility coefficients (see Eq. (\ref{Eq:convo})), without the need of a bias correction. 

The Fourier spot volatility estimator has been shown to achieve the optimal rate of convergence in the presence of noise, see \cite{mancino2022asymptotic}\footnote{This result is trivially extended to the Fourier spot co-volatility estimator if prices are sampled on a synchronous grid, as it is the case in the empirical application of Sec. \ref{Sec:results}.}. Its finite-sample efficiency at high frequencies is supported by the studies in \cite{mancino2015fourier,mancino2022asymptotic,mariotti2023fromzero}. 
 
The Fourier method also allows us to estimate the co-volatility of volatility path \citep{manbar2010}. It is worth noting that the Fourier spot co-volatility of volatility estimator does not require the pre-estimation of the spot volatility path but only the computation of integrated quantities, namely the coefficients of the co-volatility. This feature enhances the finite-sample efficiency of the Fourier methodology, compared to alternative approaches that require the pre-estimation of the spot volatility path (see \cite{tos2022jfec}). The consistency of the Fourier spot volatility of volatility estimator with noisy prices is proven in \cite{mancino2022asymptotic}. Its finite-sample performance at high frequencies is studied in \cite{Toscano}.

 \subsection{Fourier estimator of the  volatility and co-volatility}

Without loss of generality, consider the case with just two assets, that is $N=2$.  For any $i=1,2$, assume that the log-price process $p_i$ is observable on the grid $\mathcal{D}_n:=\{0=t_{0,n}<t_{1,n} \ldots <t_{n-1,n}< T=t_{n,n}\}$ such that $max_{s \in \{1,\ldots,n \}} |t_{s,n} -t_{s-1,n}| \to 0$ as $n \to \infty$\footnote{For simplicity of the exposition, we assume that two price series are synchronous, that is, are observed on the same grid $\mathcal{D}_n$, containing $n+1$ points.}. 

For $|k| \leq N_{c}$, an estimator of the $k$-th Fourier coefficient of the co-volatility is given by the convolution
  \begin{equation} \label{Eq:convo}
  c_{k}\left(C_{n, N_{c}}\right)=\frac{T}{2 N_{c}+1} \sum_{|l| \leq N_{c}} c_{l}\left(d p_{1,n}\right) c_{k-l}\left(d p_{2,n}\right),
 \end{equation}
\noindent where, for any integer $k$ such that $|k| \leq 2 N_{c}$, $c_{k}\left(d p_{i,n}\right)$ is the $k$-th  discrete  Fourier coefficient of the log-return of the $i$-th asset, namely
\begin{equation*} 
c_{k}\left(d p_{i,n}\right):=\frac{1}{T} \sum_{l=0}^{ {n}-1} e^{-\mathrm{i} k \frac{2\pi}{T} t_{l,n}} \left( p_{i}(t_{l+1,n}) -p_{i}(t_{l,n})\right), \quad i=1,2.
\end{equation*}
Similarly, for $|k| \leq N_{v_i}$, an estimator of the $k$-th Fourier coefficient of the volatility of the $i$-th asset is given by
  \begin{equation*} 
 c_{k}\left(V_{n, N_{v_i}}\right)=\frac{T}{2 N_{v_i}+1} \sum_{|l| \leq N_{v_i}} c_{l}\left(d p_{i,n}\right) c_{k-l}\left(d p_{i,n}\right), \quad i=1,2.
 \end{equation*}
Once the Fourier coefficients of the co-volatility and the individual volatilities have been
computed, applying the Fourier-Fej\'er inversion formula yields estimators of the co-volatility and volatility paths. Specifically, the Fourier estimators of the spot co-volatility and volatility at time  $\tau_b \in\mathcal{T}$ are defined as
\begin{equation*} 
\widehat{C}_{12,b}=\sum_{|k|<M_{c}}\left(1-\frac{|k|}{M_{c}}\right) c_{k}\left(C_{n, N_{c}}\right) e^{\mathrm{i} k \frac{2\pi}{T}\tau_b},
\end{equation*}
and
\begin{equation*} 
\widehat{V}_{i,b}=\sum_{|k|<M_{v_i}}\left(1-\frac{|k|}{M_{v_i}}\right) c_{k}\left(V_{n, N_{v_i}}\right) e^{\mathrm{i} k \frac{2\pi}{T}\tau_b}, \quad i=1,2,
\end{equation*}
where $M_{c} < N_c$ and $M_{v_i} < N_{v_i}$.  

\bigskip

 \subsection{Fourier estimator  of the   volatility of volatility and the co-volatility of volatility}

The knowledge of the Fourier coefficients of the latent spot co-volatility and volatility allows treating the latter as observable processes and iterating the procedure for computing the Fourier coefficients of the co-volatility of volatility and the volatility of volatility. Estimators of, respectively, the $k$-th Fourier coefficient of the co-volatility of volatility and the $k$-th Fourier coefficient of the volatility of volatility are defined as
\begin{equation*} 
c_{k}\left(\widetilde C_{n, N_{c}, S_{c}}\right)=\frac{T}{2  S_{c}+1} \sum_{|l| \leq  S_{c}}  l  (l-k) c_{l}\left(C_{n, N_{c}}\right)  c_{k-l}\left(C_{n, N_{c}}\right),  
\end{equation*}
and
\begin{equation*} 
c_{k}\left(\widetilde V_{n, N_{v_i}, S_{v_i}}\right)=\frac{T}{2  S_{v_i}+1} \sum_{|l| \leq  S_{v_i}}  l  (l-k) c_{l}\left(V_{n, N_{v_i}}\right)  c_{k-l}\left(V_{n, N_{v_i}}\right), \quad i=1,2,   
\end{equation*}
\noindent where $S_c < N_c$ and $S_{v_i}<N_{v_i}$.
The Fourier estimators of the spot co-volatility of volatility and volatility of volatility at time  $\tau_b \in\mathcal{T}$ are then defined as  
\begin{equation*} 
\widehat{\widetilde  C}_{12,b} =\sum_{|k|<L_{c}}\left(1-\frac{|k|}{L_{c}}\right)  c_{k}\left(\widetilde C_{n, N_{c}, S_{c}}\right) e^{\mathrm{i} k \frac{2\pi}{T}\tau_b},
\end{equation*} 
and
\begin{equation*} 
\widehat{\widetilde V}_{i,b} =\sum_{|k|<L_{v_i}}\left(1-\frac{|k|}{L_{v_i}}\right)   c_{k}\left(\widetilde V_{n, N_{v_i}, S_{v_i}}\right)  e^{\mathrm{i} k \frac{2\pi}{T}\tau_b}, \quad i=1,2,
\end{equation*}
\noindent where $L_c < S_c$ and $L_{v_i}<S_{v_i}$.

\section{Empirical study: an application to the universe of DJIA stocks}\label{Sec:results}

In this Section, we employ SpotV2Net for predicting the intraday spot volatilities of the 30 components of the Dow Jones Industrial Average (DJIA) index. After briefly describing the dataset and the estimation procedure for obtaining univariate and multivariate spot volatility and volatility of volatility time series from high-frequency prices, we evaluate the accuracy of the SpotV2Net model in producing single-step and multi-step forecasts.  In the case of single-step forecasts, we employ GNNExplainer \citep{ying2019gnnexplainer} to obtain insight into the most influential subgraphs related to a specific node's prediction.

\subsection{Data preparation and estimation of univariate and multivariate spot volatility and the volatility of volatility series}\label{Sec:data_est}

We collect the raw price data for the 30 components of the DJIA index from the TAQ - Millisecond Consolidated Trades database, which we access through the Wharton Research Data Services (WRDS). 
The dataset covers the period from June 1\textsuperscript{st}, 2020, to May 10\textsuperscript{th}, 2023, corresponding to 737 days.

For the implementation of Fourier estimator, we proceed as follows. Starting from tick data, we filter the transactions to include only those executed on the New York Stock Exchange (NYSE) and exclude transactions that occurred outside the operating market hours from 9:30 a.m. to 4:00 p.m. Then, we resample the tick data to the one-second frequency, selecting the price associated with the transaction closest to each second's onset. The resulting sample size employed in the implementation of the estimators is $n=23400$ per day. Note that we measure time in days and set T equal to 1. The cutting frequencies $N_{v_i}$, $M_{v_i}$, $S_{v_i}$, $L_{v_i}$ (for the $i$-th asset) and $N_c$,  $M_c$, $S_c$ and $L_c$ (for each pair of assets) are selected using the guidance from the paper by \cite{SanTos24}, which introduces a MATLAB library for the implementation of Fourier estimators\footnote{The Fourier method is not consistent in the presence of jumps. To remove the effect of jumps from the estimation, we identify 1-second returns whose absolute value is larger than a given threshold $\vartheta_n$ and replace them with zeros. Letting $\vartheta_n=\beta \left( T/n\right)^\alpha$, the constants that determine the threshold are selected using numerical simulations. As a result, we find that suitable choices are $\beta=0.5$ and $\alpha=0.5$.}.  Univariate and multivariate spot volatility and volatility of volatility estimates are obtained on the 30-minute grid.  
 
As the output of the univariate estimation process, for each of the $N=30$ DJIA components, we obtain a time series that contains $10318$\footnote{The dataset covers $737$ days. For each day (i.e., $6.5$ hours) we obtain $14$ estimates on the 30-minute grid (note that the last estimate was performed at 3:59 p.m. instead of 4:00 p.m., to reduce the effects of periodicity on Fourier estimates). Hence the total of $737\times 14=10318$ estimates.} spot volatility values and a time series with the same length that contains spot volatility of volatility values. Furthermore, the output of the multivariate estimation process includes $\frac{1}{2} N\times (N-1)$ time series with length $10318$ for the spot co-volatilities and other $\frac{1}{2} N\times (N-1)$ time series of the same length for the spot co-volatilities of volatility.  

Fourier estimates time series are partitioned into three sets: train, validation and test. Tab. \ref{tab:data_splits} outlines the start and end dates of each set, along with the proportion of observations included in each set. For each set, every $30$ minutes we set up a static fully connected graph which includes contemporaneous and lagged Fourier estimates as node and edge features, according to the structure outlined in Sec. \ref{Sec:model}.

\begin{table}[h!]
\centering
\begin{tabular}{cccc}
\hline
& \textbf{Train} & \textbf{Validation} & \textbf{Test} \\
\hline
\textbf{Start} & 06/01/2020 & 07/21/2022 & 10/15/2022 \\

\textbf{End} & 07/20/2022 & 10/14/2022 & 05/10/2023 \\

\textbf{\# of 30 min obs} & 7518 & 840 & 1960 \\

\textbf{Proportion} & 73\% & 8\% & 19\% \\
\hline
\end{tabular}
\caption{The table delineates the periods of the dataset allocated for the train, validation and test sets, with the corresponding start and end dates. It also details the number of 30-minute observations employed in each period and their proportions within the dataset.}
\label{tab:data_splits}
\end{table}
\FloatBarrier
 
\subsection{Single-step forecast}\label{sec:single_step_forecast}

In this Section, we present the results of the application of SpotV2Net for predicting the 30-minute-ahead spot volatility of the entire cross-section of DJIA constituents.  We train the model using the data-splitting strategy detailed in Tab. \ref{tab:data_splits} and use the validation set to fine-tune the hyperparameter choices. The test set is kept completely separate and is used for out-of-sample forecast evaluation. The selection of hyperparameters is documented in App. \ref{Sec:appendix_hyperparameters}.

Note that our focus is on predicting volatilities, i.e., the diagonal elements of the covariance matrix, rather than the entire covariance matrix. While the model architecture is inherently flexible and could be adapted to forecast the entire covariance matrix, we leave this extension to future research. This decision is motivated by the challenge posed by the curse of dimensionality, which causes the number of parameters in the model to rapidly increase with the cross-sectional dimension, with negative consequences regarding computational costs.

We compare the forecasting accuracy of SpotV2Net with that of three alternative models, detailed in App. \ref{sec:appendix_alternativemodels}: the panel HAR-Spot model, the XGB model and LSTM model. We also compare the forecasting accuracy of SpotV2Net with that of a modified version of the latter, which does not accept edge features as input but only node features. We refer to it as SpotV2Net with no edge features, shortened to SpotV2Net-NE.   Tab. \ref{tab:results} shows aggregate values of the Mean Squared Error (MSE), that is, 
\begin{equation}\label{eq:aggMSE}
 \frac{1}{T_{\text{test}}} \sum_{j=1}^{T_{\text{test}}} \frac{1}{N} \sum_{i=1}^{N} \left( \widehat{V}_{i,j} - \widehat{V}^*_{i,j} \right)^2,
\end{equation}
and the QLIKE, that is,
\begin{equation}\label{eq:aggQLIKE}
{\color{black} \frac{1}{T_{\text{test}}} \sum_{j=1}^{T_{\text{test}}}  \frac{1}{N} \sum_{i=1}^{N} \left[ \frac{ \widehat{V}_{i,j}}{ \widehat{V}^*_{i,j}} - \log \frac{ \widehat{V}_{i,j}}{\widehat{V}^*_{i,j}} - 1 \right]},
 \end{equation}
where $\widehat{V}^*_{i,j}$ denotes the forecast, at time $\tau_{j-1}$, of the spot volatility of $i$-th asset at time $\tau_j$,  and $T_{\text{test}}$ is the length of the test period as indicated in Tab. \ref{tab:data_splits}. When the validation period is considered, $T_{\text{test}}$ is replaced by $T_{\text{val}}$, which indicates the length of the validation period.

\begin{table}[h]
\centering
\begin{tabular}{lcccc}
\toprule
& \multicolumn{2}{c}{Validation} & \multicolumn{2}{c}{Test} \\
\cmidrule(lr){2-3} \cmidrule(lr){4-5}
Single-step model & MSE & QLIKE & MSE & QLIKE \\
\midrule
% ARFIMA & 3.433e-08 & 0.238 & 5.487e-08 & 0.379 \\[1.0ex]
HAR-Spot & 3.797e-08 & 0.263 & 5.089e-08 & 0.343 \\[1.0ex]
XGB & 3.148e-08 & 0.217 & 6.232e-08 & 0.429 \\[1.0ex]
LSTM & 2.629e-08 & 0.184 & 5.198e-08 & 0.359 \\[1.0ex]
SpotV2Net-NE & 2.513e-08 & 0.173 & 5.115e-08 & 0.354 \\[1.0ex]
%SpotV2Net & 2.299e-08* & 0.159* & 4.885e-08* & 0.337* \\[1.0ex]
{SpotV2Net} & 2.079e-08* & 0.137* & 4.273e-08* & 0.286* \\[1.0ex]
\bottomrule
\end{tabular}
\caption{Aggregate values of MSE and QLIKE in Eqns. (\ref{eq:aggMSE}) and (\ref{eq:aggQLIKE}) for the SpotV2Net model and the alternative models during the validation and test phases. An asterisk (*) indicates models that are included in the MCS at the 5\% significance level.}
\label{tab:results}
\end{table}
\FloatBarrier

Aggregate performance measures in Tab. \ref{tab:results} suggest that SpotV2Net outperforms the HAR-Spot model and the other ML models considered, namely XGB and LSTM, both during the validation and test periods. 
Besides, while the LSTM and XGB models represent an improvement in forecasting accuracy compared to the HAR-Spot model in the validation period, the latter shows a better performance, compared to LSTM and XGB, in the test period. SpotV2Net-NE is less accurate than SpotV2Net in both the validation and the test phases. 
Moreover, it is worth noting that SpotV2Net-NE still outperforms HAR-Spot and the other ML methods in the validation period. Instead, in the test period, SpotV2Net-NE performs worse than HAR-Spot, but still better than LSTM and XGB. These findings underline the crucial role of edge features (that is, the values of the spot co-volatility of volatility matrix) in capturing spillover effects and improving forecasting accuracy. 
Note that the improved performance by all models in the validation period is due to the fact that the validation employs data for fine-tuning model hyperparameters.

Tab. \ref{tab:results} also reports the results of the Model Confidence Set (MCS) \citep{hansen2011model}, which is applied column-wise to identify a subset of models that demonstrate superior performance with a 95\% level of confidence. \textcolor{black}{We implemented the MCS with 5000 bootstrap replications, using the range statistic to compute the $p$-values.}  
\textcolor{black}{The results show that the subset of superior models contains only SpotV2Net,
indicating that the latter yields superior predictive performance, compared to the other models considered, with a level of confidence of 95\%.}

Moreover, we also employ the Diebold-Mariano (DM) test \citep{diebold2002comparing} to compare the forecasting accuracy between models at the 5\% significance level. 
The DM test statistics are provided in Tabs. \ref{Tab:DB_val_single} and \ref{Tab:DB_test_single} for the validation and the test set, respectively. The results of the DM test suggest that the gain in forecasting accuracy provided by SpotV2Net is statistically significant.

\begin{table}[!h]
    \centering
    \small
   \textbf{Validation period (MSE)} 
   
    \begin{tabular}{lcccc}
        \toprule
         &  XGB & LSTM & SpotV2Net-NE & SpotV2Net \\
        \midrule
        HAR-Spot   & 6.32* & 40.23* & 42.12* & 45.18* \\
        XGB  & -- & 5.50* & 6.16* & 25.94* \\
        LSTM  & -- & -- & 2.34* & 9.08* \\
        SpotV2Net-NE   & -- & -- & -- & 6.12* \\
        \bottomrule
    \end{tabular}

    \textbf{Validation period (QLIKE)} 
    
        \begin{tabular}{lcccc}
        \toprule
          &  XGB & LSTM & SpotV2Net-NE & SpotV2Net \\
        \midrule
        HAR-Spot   & 13.11* & 21.05* & 23.45* & 28.14* \\
        XGB & --   & 12.32* & 13.06* & 20.06* \\
        LSTM & --   & -- & 5.67* & 12.43* \\
        SpotV2Net-NE   & -- & -- & -- & 8.91* \\
        \bottomrule
    \end{tabular}
    \caption{Comparison of single-step forecasting accuracy using the DM test with MSE (top) and QLIKE (bottom) losses in the validation period. Positive numbers indicate the column model outperforms the row model. Superscript * denotes the significance levels of 5\%.}
    \label{Tab:DB_val_single}
\end{table}
\FloatBarrier

\begin{table}[!h]
    \centering
    \small
 \textbf{Test period (MSE)}  
 
    \begin{tabular}{lcccc}
        \toprule
 
          & XGB & LSTM & SpotV2Net-NE & SpotV2Net \\
        \midrule
        HAR-Spot &   -42.69* & -1.11 & -0.38 & 6.71* \\
        XGB &   -- & 37.89* & 43.25* & 52.12* \\
        LSTM &   -- & -- & 0.71 & 11.87* \\
        SpotV2Net-NE &   -- & -- & -- & 7.23* \\
            \bottomrule
        \end{tabular}

    \textbf{Test period (QLIKE)} 
    
    \begin{tabular}{lcccc}
        \toprule
          & XGB & LSTM & SpotV2Net-NE & SpotV2Net \\
        \midrule
        HAR-Spot   & -44.34* & -3.17* & -2.41* & 6.57* \\
        XGB  & -- & 22.98* & 26.15* & 32.54* \\
        LSTM   & -- & -- & 0.55 & 10.62* \\
        SpotV2Net-NE   & -- & -- & -- & 7.05* \\
             \bottomrule
        \end{tabular}
    \caption{Comparison of single-step forecasting accuracy using the DM test with MSE (top) and QLIKE (bottom) losses in the test period. Positive numbers indicate the column model outperforms the row model. Superscript * denotes the significance levels of 5\%.}
    \label{Tab:DB_test_single}
\end{table}

\subsection{Model interpretation}

The complexity of the GAT architecture, on which SpotV2Net is based, suggests using specific methodologies to enhance interpretability. Indeed, model interpretation can be particularly important for applications in the financial domain. In particular, it may be relevant to pinpoint which nodes (i.e., which asset volatilities) are the most influential in predicting the volatility of a specific node, thereby obtaining insight into spillover effects and volatility transmission channels captured by SpotV2Net.

For this purpose, we employ GNNExplainer \citep{ying2019gnnexplainer}, a model-agnostic approach that provides interpretable explanations for predictions of GNN-based models. 
Specifically, GNNExplainer allows the identification of the subgraph\footnote{A subgraph is defined as a graph consisting of a subset of the nodes and edges of the original graph.} deemed the most influential for the model's prediction of a specific node.

We briefly illustrate how GNNExplainer works. Consider a trained GNN-based model, denoted by $\mathcal S$ and let $\mathcal G$ denote the input graph of  $\mathcal S$, composed of $N$ nodes. Further, let $\hat y_{i,\mathcal{S}(\mathcal{G})}$ denote prediction produced by  $\mathcal S$ for the target variable $y_i$ associated with the $i$-th node of $\mathcal{G}$. For brevity, we use the shorthand notation 
$\hat{y_i}:=\hat y_{i,\mathcal{S}(\mathcal{G})}$. In the case of SpotV2Net, $\hat y_i$ and $y_i$ will be, respectively, the spot volatility prediction and the actual spot volatility value
associated with the $i$-th node of the input graph described in Sec. \ref{Sec:model}. 

For a given node $i$ of $\mathcal{G}$, GNNExplainer aims to select the subgraph $\mathcal{G}^*$, composed of $N^* < N$ nodes, that maximizes the mutual information (MI) between the prediction $\hat y_{i}$ and the prediction $\hat{y}^*_i:=\hat{y}_{i,\mathcal{S}(\mathcal{G}^*)}$, which indicates the prediction produced by $\mathcal{S}$ for node $i$ when its input is restricted to the subgraph $\mathcal{G}^*$.

Let $p_i:=p_{i,\mathcal{S}(\mathcal{G})}$ and $p^*_i:=p^*_{i,\mathcal{S}(\mathcal{G}^*)}$ denote, respectively, the density of the forecasts $\hat y_{i}$ and $\hat{y}^*_{i}$. The corresponding supports are denoted by $\mathcal{Y} \subseteq  \mathbb{R}$ and $\mathcal{Y}^{*} \subseteq  \mathbb{R}$. The optimization problem solved by GNNExplainer reads as
\begin{equation}\label{Eq:entropy_max}
\max_{\mathcal{G}^*}  MI_{i, \mathcal{G}, \mathcal{G^*}} =  h(\hat y_{i}) - h(\hat y^*_{i}),
\end{equation}
where $h(\hat y_{i})=- \int_{\mathcal Y} p_{i}(x) \,\, \mathrm{log} \, p_{i(x)} dx $ and $ h(\hat y^*_{i})=- \int_{\mathcal Y^*} p^*_{i}(x) \, \,\mathrm{log} \, p^*_{i}(x) dx $ denote, respectively, the entropy of $\hat{y}_i$ and $\hat{y}^*_i$. As the entropy of a forecast represents the level of uncertainty related to the forecast's outcomes,  $ MI_{i, \mathcal{G}, \mathcal{G^*}}$ in Eq. (\ref{Eq:entropy_max}) quantifies the difference in such uncertainty between the case when the input graph of $\mathcal{S}$ is the full graph $\mathcal{G}$ and the case when the input graph of $\mathcal{S}$ is restricted to $\mathcal{G}^*$. The larger the difference in Eq. (\ref{Eq:entropy_max}), the more the subgraph $\mathcal{G}^*$ is deemed influential in predicting the target $y_i$. In other words, if removing the uncertainty ascribed to subgraph $\mathcal{G}^*$ maximizes the uncertainty related to the forecast $\hat y_i$, then the subgraph $\mathcal{G}^*$ is deemed a relevant $counterfactual$ explanation of the forecast $\hat y_i$ (see Section 4.1 of \cite{ying2019gnnexplainer} for further details).

Since $h(\hat{y}_i)$ is independent of $\mathcal{G}^*$, the maximization problem in Eq. (\ref{Eq:entropy_max}) is equivalent to the minimization problem 
\begin{equation*}
\min_{\mathcal{G}^*}   h(\hat y^*_{i}).
\end{equation*}

For the interpretation of the single-step forecasts produced by SpotV2Net  for the spot volatilities of the DIJA constituents, we implement GNNExplainer for each of the $N=30$ nodes (i.e., asset spot volatilities) of the input graph, with the aim of extrapolating, for each node, a subgraph composed of the $N^*=5$ nodes representing the most influential nodes for predicting the node's spot volatility. \textcolor{black}{
%Our use of the GNNExplainer 
Note that we use GNNEXPLAINER to identify the most influential subgraph, not to isolate the most influential individual features of specific nodes. %Since each node contains a set of features, 
This approach provides a global perspective on the relevance of a particular subgraph based on the entire information set embedded in its nodes and edges. This way, GNNExplainer complements traditional econometric methods, which typically emphasize the statistical significance of the contribution of individual features.} The implementation is repeated at each time step\footnote{For the implementation, we use the Python package available at the link \href{https://pytorch-geometric.readthedocs.io/en/latest/generated/torch_geometric.explain.algorithm.GNNExplainer.html\#torch_geometric.explain.algorithm.GNNExplainer}{GNNExplainer}. Details about the optimization algorithm are provided in Section 4 of \cite{ying2019gnnexplainer}.}.

\begin{figure}
    \centering
    \begin{minipage}{0.48\textwidth}
        \centering
        \includegraphics[width=\textwidth]{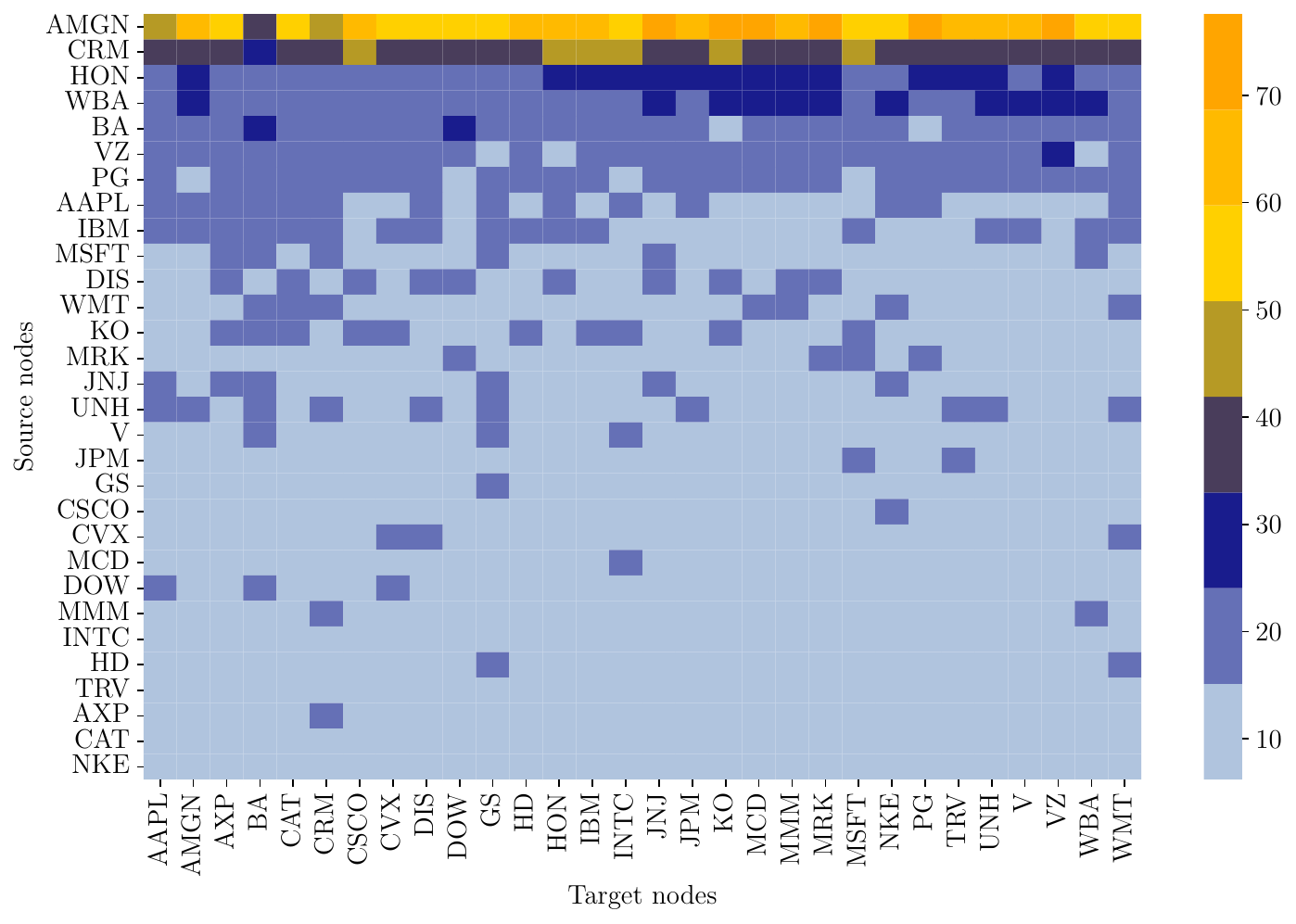}
    \end{minipage}
    \hfill
    \begin{minipage}{0.48\textwidth}
        \centering
        \includegraphics[width=\textwidth]{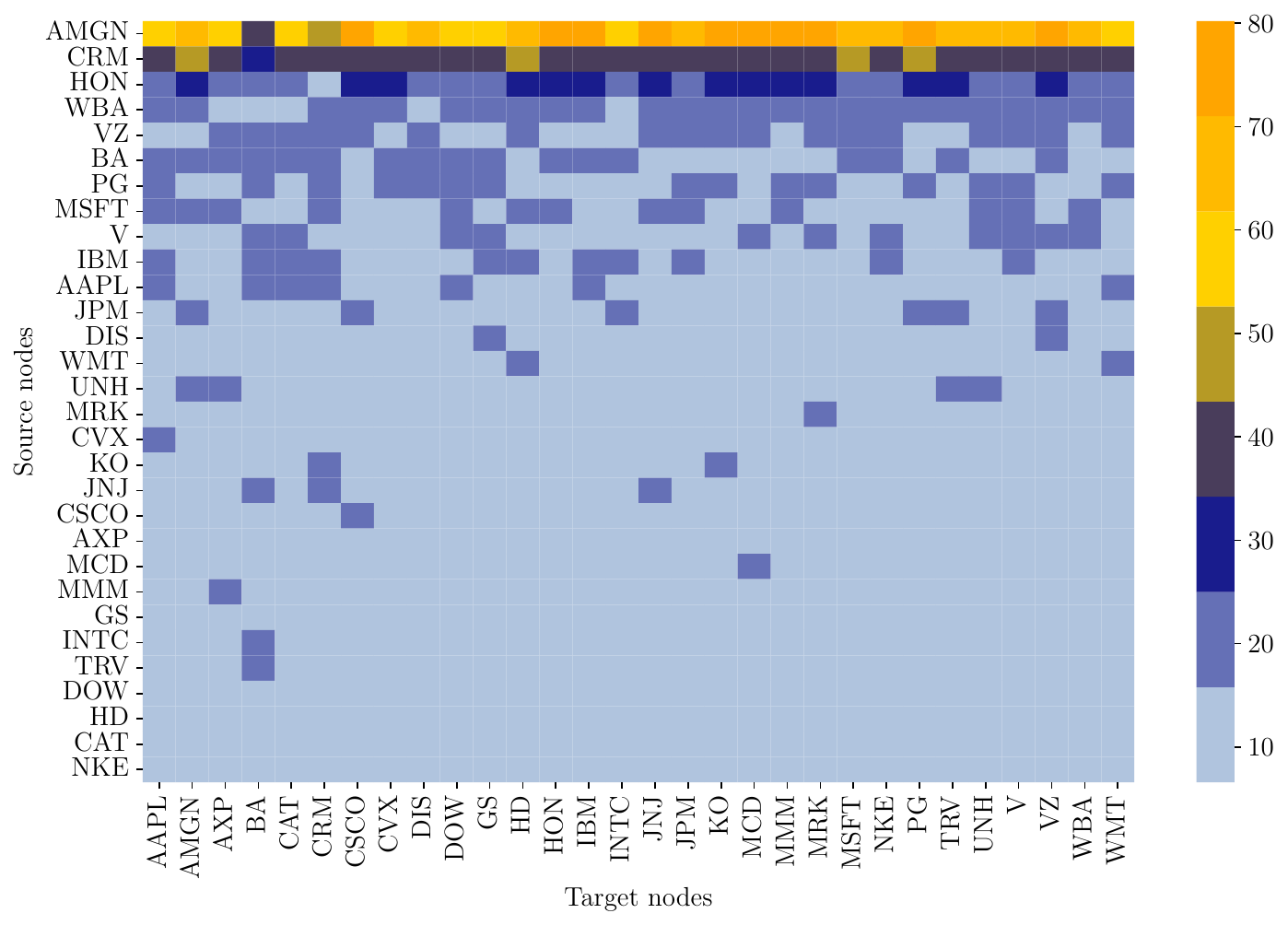}
    \end{minipage}
     \caption{Heatmaps assigning to nodes on the y-axis ('source nodes')  the frequency (in percentage terms) by which GNNExplainer includes them in the most influential subgraphs (with $N^*=5$ nodes) for predicting the spot volatility of nodes on the x-axis ('target nodes'), in the validation set (left panel) and test set (right panel). For each target node, GNNExplainer is implemented at every timestamp (i.e., single-step forecasts). The frequency of inclusion is computed by counting the number of timestamps (i.e., single-step forecasts) for which a node is retained in the subgraph of the most influential nodes.}
    \label{fig:gnn_explainer}
\end{figure}

The results of the implementation of GNNExplainer are summarized in Fig. \ref{fig:gnn_explainer} using two heatmaps, one for the validation set and one for the test set, which illustrate the frequency (in percentage terms) of the inclusion of nodes into the subgraphs of the most influential nodes for predicting the spot volatility of a given asset.

\FloatBarrier

 Fig. \ref{fig:gnn_explainer} shows that Amgen (AMGN), Salesforce (CRM) and Honeywell (HON) are deemed the three most influential nodes for predicting the intraday spot volatility of all stocks in the dataset, in both the validation and test phases. Interestingly, these three companies became part of the DJIA index at the end of August 2020, near the beginning of the dataset, which starts in June 2020. More specifically, GNNExplainer suggests that univariate and multivariate volatility and volatility of volatility of the three new index constituents - included in node and edge feature vectors, respectively - are often influential in predicting the intraday spot volatility of all other assets during the period covered by our study. In particular, note that AMGN is included in the most influential subgraphs for other nodes with a frequency between approximately 40\% and 80\%, in both the validation and test sets. CRM and HON follow with frequencies that range approximately between 40\% and 50\% (for CRM) and 10\% and 30\% (for HON).

 These findings may be explained by the fact that the training period of SpotV2Net, which ranges tentatively from mid-2020 to mid-2022, corresponds to eventful contingencies for these three companies. At that time,  Amgen was conducting its research on the COVID-19 vaccine, Salesforce was experiencing an increase in the demand for cloud platforms, due to the rise in remote work caused by the pandemic and the related lockdowns, and Honeywell was converting part of its manufacturing facilities to produce protective equipment for healthcare workers to reduce contagion risks. Furthermore, it is worth noting that the features of Verizon (VZ) and Walgreens (WBA) are also recognized as influential for nodes' predictions by GNNExplainer with frequencies that range approximately between 10\% and 30\%. This may be explained by the fact that their businesses (telecommunications and retail pharmacy, respectively) were closely tied to pandemic-related contingencies. 
 Overall, the GNNExplainer-based interpretation of the relationships learned by SpotV2Net during the training period may suggest the presence of a drag effect, coming from the volatility of certain DJIA index components related to key economic sectors for the observed period.

\subsection{Multi-step forecast}

Accurate multi-step forecasting can be advantageous for high-frequency traders, who wish not only to predict volatilities at the immediate next time step but also to extrapolate entire trajectories, enabling them to plan their operations with a broader, daily, temporal perspective.
Accordingly, in this Section, we extend the scope of application of the SpotV2Net model to obtain multiple-step-ahead forecasts. Specifically,  we adapt the model's architecture to predict the entire daily volatility function on the 30-minute grid, which corresponds to 14 future observations for a 6.5-hour long trading day \footnote{In principle, it is possible to adapt SpotV2Net's architecture to produce multi-step forecasts at any point in time. For simplicity, we restrict our study to multi-step forecasts obtained at the end of a trading day for the entire next trading day.}. 
Note that multi-step forecasts are all conditional on the information available at the end of the previous trading day (i.e., node and edge features of the graph at the end of the previous trading day). In other words, letting $\tau_0$ denote the end of the previous day, the multi-step version of SpotV2Net does not use as input the predictions at times $\tau_{1},...,\tau_{h-1}$ to obtain the prediction at time $\tau_{h}$, $1<h\le 14$. Instead, the model produces, at $\tau_0$, a unique joint forecast of the values of the spot volatility function on the time grid $\tau_{1},...,\tau_{14}$, with mesh size equal to $30$ minutes. \textcolor{black}{We refer to this type of multi-step forecast as functional forecast}. We evaluate the performance of SpotV2Net in producing such type of forecast against the same alternative models considered in Sec. \ref{sec:single_step_forecast}.\footnote{The ability to produce a \textcolor{black}{functional} forecast is characteristic of deep learning models, like SpotV2Net and the LSTM model. However, such an ability is not shared by the HAR-Spot model and the XGB model. Indeed, to obtain multi-step forecasts with these models, one has to resort to a recursive approach, where the one-step ahead forecast for time $\tau_j$ is employed as an input to obtain the one-step ahead forecast for time $\tau_{j+1}$.}.

\bigskip 

  \textcolor{black}{To evaluate functional forecast accuracy, we use $global$ loss functions. Specifically, we employ the average MSE of the multi-step prediction, that is,}
\begin{equation}\label{eq:aggMSE2}
  \frac{1}{D_{\text{test}}} \sum_{j=1}^{D_{\text{test}}} \frac{1}{N} \sum_{i=1}^{N} \frac{1}{H} \sum_{h=1}^{H} \left( \widehat{V}_{i,j+h} - \widehat{V}^*_{i,j+h} \right)^2 
\end{equation}
\textcolor{black}{and the average QLIKE, that is,}
\begin{equation}\label{eq:aggQLIKE2}
\textcolor{black}{ \frac{1}{D_{\text{test}}} \sum_{j=1}^{D_{\text{test}}} \frac{1}{N} \sum_{i=1}^{N} \frac{1}{H} \sum_{h=1}^{H} \left[ \frac{ \widehat{V}_{i,j+h} }{ \widehat{V}^*_{i,j+h} } -  \log \frac{\widehat{V}_{i,j+h}}{\widehat{V}^*_{i,j+h}} - 1 \right],  } 
\end{equation}
where $H=14$ denotes the number of intraday 30-minute forecast steps and $D_{test}$ indicates the number of days in the test set. When the validation period is considered, $D_{\text{test}}$ is replaced by $D_{\text{val}}$, which indicates the number of days in the validation period.

Tab. \ref{tab:results_multi} summarizes the multi-step forecasting performance of the models employed in our study, in the validation and test phases. In the validation period, the ranking is the same as for single-step forecasts, both in terms of aggregate values of \textcolor{black}{average} MSE and QLIKE. Instead, in the test period, the HAR-Spot model has a worse ranking than in the single-step exercise, as it outperforms only the XGB model. This finding is in line with the results by \cite{christensen2023machine}, which suggest that the relative forecasting accuracy of HAR-type models may deteriorate at longer forecasting horizons, compared to that of  ML models. Tab. \ref{tab:results_multi} reports also the results of the application of the MCS at the \textcolor{black}{95\%  confidence level}. The outcome is analogous to the one obtained in the single-step case and supports the superior forecasting performance of SpotV2Net also in the multi-step exercise.

\begin{figure}[h]
    \centering
    % First row of subplots (Validation period)
    \begin{minipage}[b]{0.23\textwidth}
        \includegraphics[width=\textwidth]{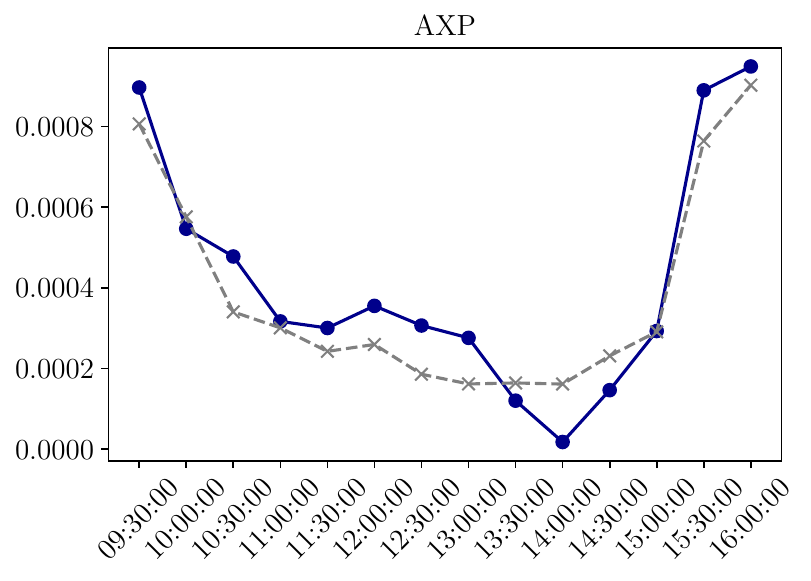}
    \end{minipage}
    \hspace{4pt} % Adjust spacing between subplots
    \begin{minipage}[b]{0.23\textwidth}
        \includegraphics[width=\textwidth]{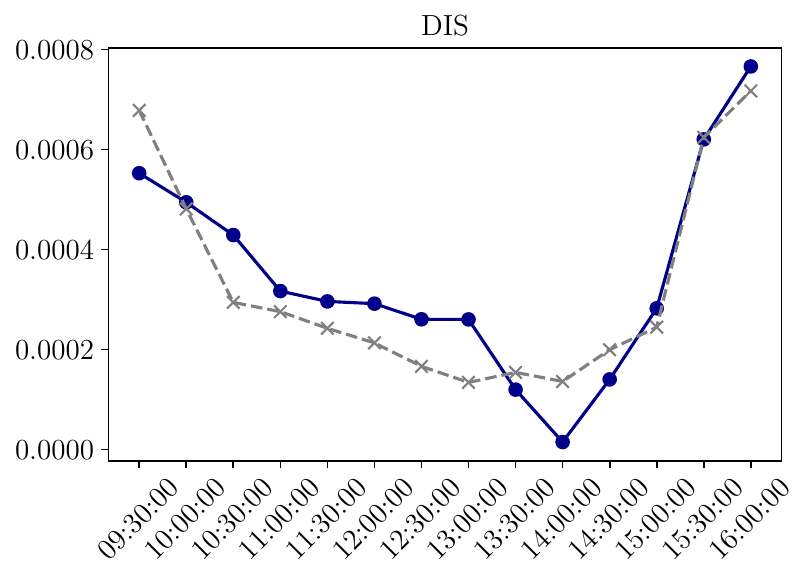}
    \end{minipage}
    \hspace{4pt} % Adjust spacing between subplots
    \begin{minipage}[b]{0.23\textwidth}
        \includegraphics[width=\textwidth]{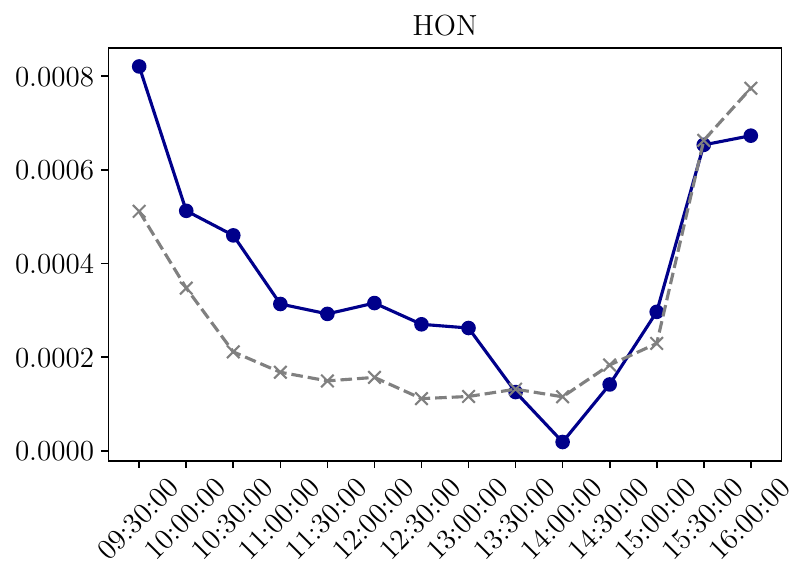}
    \end{minipage}
    \hspace{4pt} % Adjust spacing between subplots
    \begin{minipage}[b]{0.23\textwidth}
        \includegraphics[width=\textwidth]{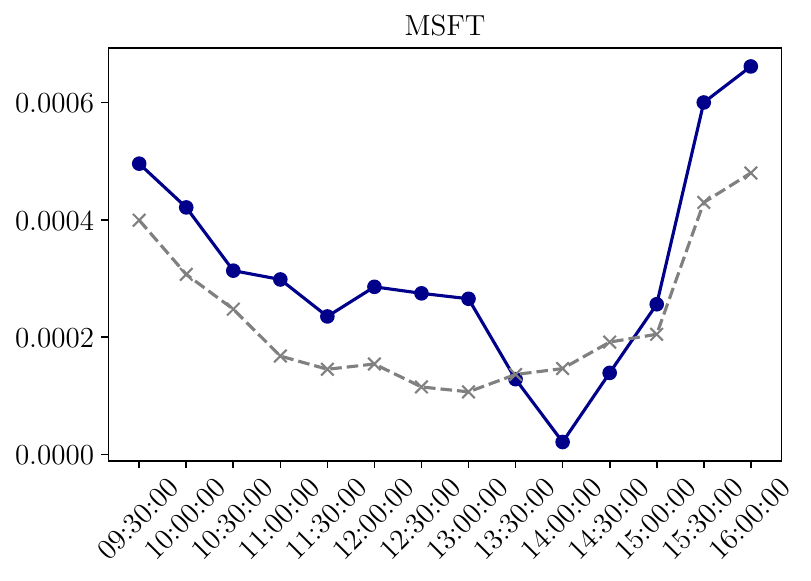}
    \end{minipage}
    
    \vspace{10pt} % Adjust vertical spacing between rows
    
    % Second row of subplots (Test period)
    \begin{minipage}[b]{0.23\textwidth}
        \includegraphics[width=\textwidth]{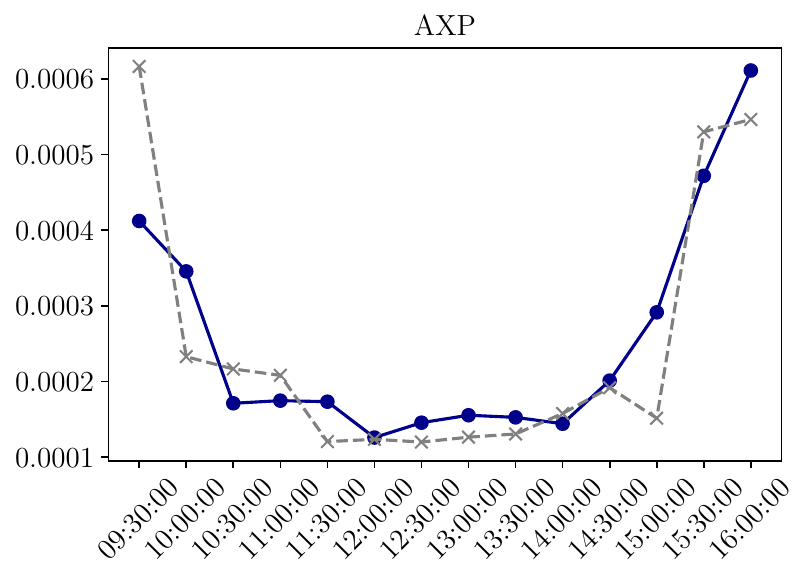}
    \end{minipage}
    \hspace{4pt} % Adjust spacing between subplots
    \begin{minipage}[b]{0.23\textwidth}
        \includegraphics[width=\textwidth]{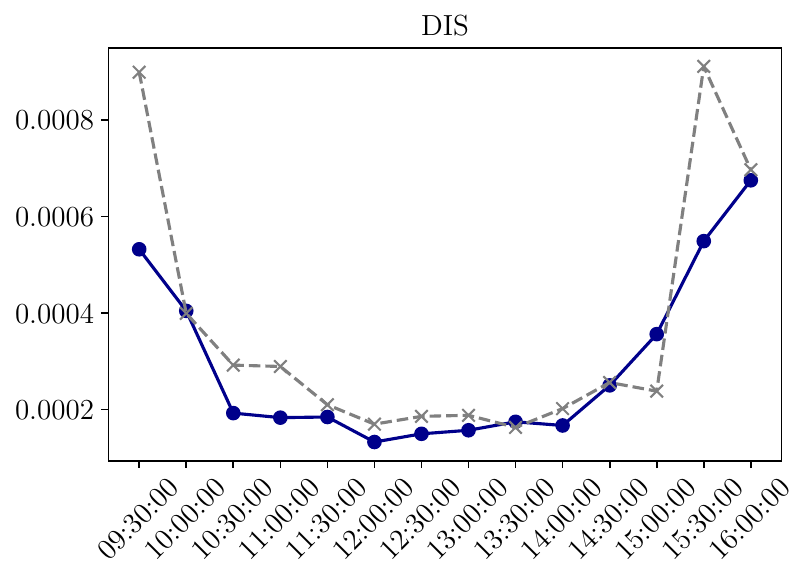}
    \end{minipage}
    \hspace{4pt} % Adjust spacing between subplots
    \begin{minipage}[b]{0.23\textwidth}
        \includegraphics[width=\textwidth]{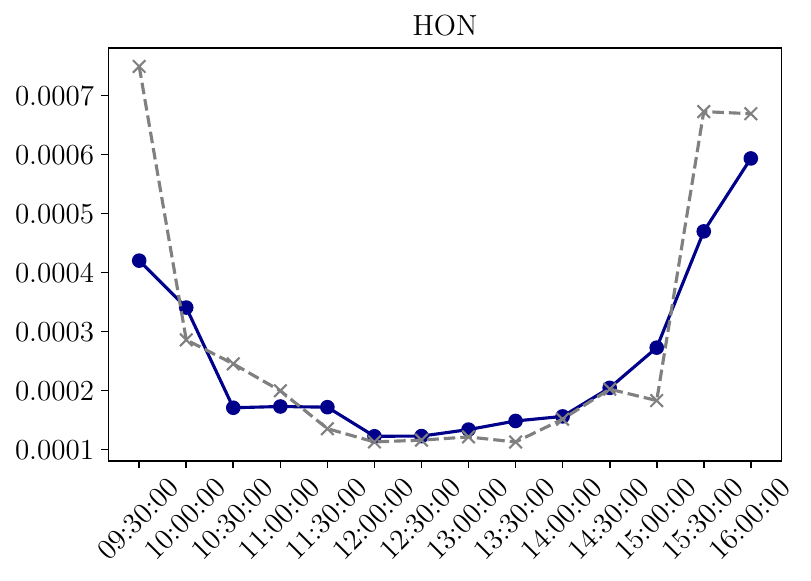}
    \end{minipage}
    \hspace{4pt} % Adjust spacing between subplots
    \begin{minipage}[b]{0.23\textwidth}
        \includegraphics[width=\textwidth]{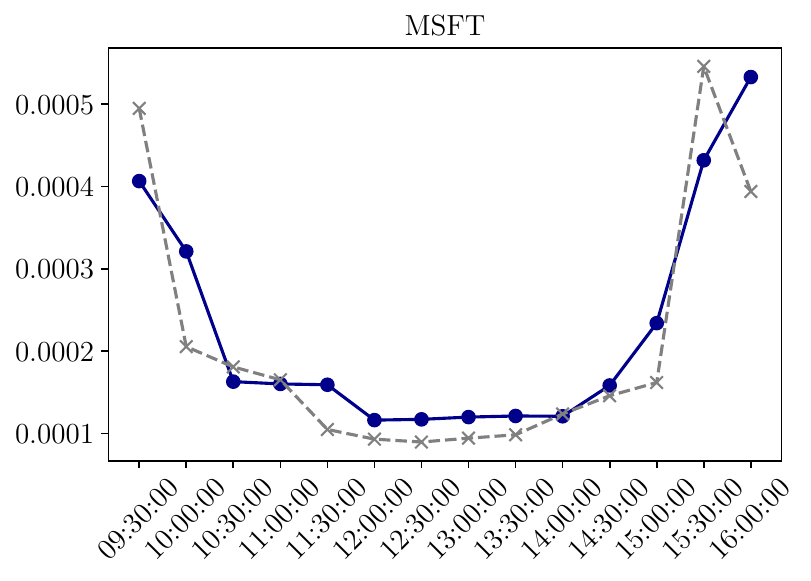}
    \end{minipage}
        \caption{\textcolor{black}{Predicted (dotted blue) and actual (reconstructed, dashed grey) spot volatility path for AXP, DIS, HON, and MSFT. The first row refers to September 19, 2022, comprised in the validation period, and the second row to February 1, 2023, comprised in the test period.}}
    \label{Fig:combined_multi}
\end{figure}

Moreover, the DM test statistics are reported in Tabs. \ref{Tab:DB_val_multi} and \ref{Tab:DB_test_multi}.  The results suggest that the gain in forecasting accuracy provided by SpotV2Net is statistically significant at the 5\% level also for multi-step forecasts. 
\textcolor{black}{Finally, to illustrate an example of the accuracy of the multi-step functional predictions produced by SpotV2Net, Fig. \ref{Fig:combined_multi} shows a comparison of the latter with the actual reconstructed spot volatility path of four companies for two days, one in the validation set and one in the test set.}

\begin{table}[h]
\centering
\begin{tabular}{lcccc}
\toprule
& \multicolumn{2}{c}{Validation} & \multicolumn{2}{c}{Test} \\
\cmidrule(lr){2-3} \cmidrule(lr){4-5}
Multi-step model &  {MSE} & {QLIKE} &  {MSE} &  {QLIKE} \\
\midrule
HAR-Spot & 4.453e-08 & 0.310 & 5.876e-08 & 0.407 \\[1.0ex]
XGB & 4.156e-08 & 0.291 & 6.943e-08 & 0.478 \\[1.0ex]
LSTM & 2.923e-08 & 0.204 & 5.467e-08 & 0.379 \\[1.0ex]
SpotV2Net-NE & 2.643e-08 & 0.180 & 5.312e-08 & 0.368  \\[1.0ex]
%SpotV2Net & 2.398e-08* & 0.165* & 4.580e-08* & 0.317* \\[1.0ex]
 SpotV2Net  & 2.145e-08* & 0.142* & 4.342e-08* & 0.294* \\[1.0ex]
\bottomrule
\end{tabular}
\caption{Aggregate values of \textcolor{black}{average} MSE and QLIKE in Eqns. (\ref{eq:aggMSE2}) and (\ref{eq:aggQLIKE2}) for the SpotV2Net model and the alternative models during the validation and test phases. An asterisk (*) indicates models that are included in the MCS at the 5\% significance level.}
\label{tab:results_multi}
\end{table}

\begin{table}[h]
    \centering
    \small
     \textbf{Validation period ({MSE})} % Title to distinguish table
    
    \begin{tabular}{lcccc}
        \toprule
          &   XGB & LSTM & SpotV2Net-NE & SpotV2Net \\
        \midrule
        HAR-Spot & 7.12* & 46.76* & 48.21* & 58.23* \\
        XGB & -- & 28.55* & 35.15* & 47.36* \\
        LSTM  & -- & -- & 5.64* & 11.98* \\
        SpotV2Net-NE  & -- & -- & -- & 8.53* \\
        \bottomrule
    \end{tabular}
    
    \hspace{0.25cm} % Space between tables
    
    \textbf{Validation period (QLIKE)}

    \begin{tabular}{lcccc}
        \toprule
           & XGB & LSTM & SpotV2Net-NE & SpotV2Net \\
        \midrule
        HAR-Spot   & 5.41* & 30.20* & 37.05* & 43.82* \\[1.0ex]
        XGB   & -- & 24.79* & 31.63* & 38.37* \\[1.0ex]
        LSTM   & -- & -- & 6.83* & 14.51* \\[1.0ex]
        SpotV2Net-NE   & -- & -- & -- & 7.09* \\[1.0ex]
        \bottomrule
    \end{tabular}
    \caption{Comparison of multi-step forecasting accuracy using the DM test with \textcolor{black}{average} MSE (top) and QLIKE (bottom) losses on the validation set. Positive numbers indicate the column model outperforms the row model. Superscript * denotes the significance levels of 5\%.}
    \label{Tab:DB_val_multi}
\end{table}

\begin{table}[h]
    \centering
    \small
  \textbf{Test period ({MSE})}  
  
    \begin{tabular}{lcccc}
        \toprule
         &    XGB & LSTM & SpotV2Net-NE & SpotV2Net \\
        \midrule
        HAR-Spot  & -25.57* & 9.80* & 13.52* & 33.16* \\[1.0ex]
        XGB & -- & 35.38* & 39.10* & 59.51* \\[1.0ex]
        LSTM  & -- & -- & 3.71*  & 24.61* \\[1.0ex]
        SpotV2Net-NE  & -- & -- & -- & 21.01* \\[1.0ex]
        \bottomrule
    \end{tabular}

    \vspace{0.25cm}  

    \textbf{Test period (QLIKE)}  
    
    \begin{tabular}{lccccc}
        \toprule
          & XGB & LSTM & SpotV2Net-NE & SpotV2Net \\
        \midrule
        HAR-Spot   & -20.23* & 7.90* & 11.15* & 28.57* \\[1.0ex]
        XGB   & -- & 28.21* & 31.34* & 46.95* \\[1.0ex]
        LSTM   & -- & -- & 3.13* & 21.34* \\[1.0ex]
        SpotV2Net-NE   & -- & -- & -- & 17.15* \\[1.0ex]
        \bottomrule
    \end{tabular}
    \caption{Comparison of multi-step forecasting accuracy using the DM test with \textcolor{black}{average} MSE (top) and QLIKE (bottom) losses on the test set. Positive numbers indicate the column model outperforms the row model. Superscript * denotes the significance levels of 5\%.}
    \label{Tab:DB_test_multi}
\end{table}
\FloatBarrier

\section{Conclusions}\label{Sec:conclusion}

This paper contributes to research on volatility forecasting by introducing SpotV2Net, a model that employs a neural network with a GAT architecture to produce intraday forecasts of multivariate spot volatilities. Specifically, SpotV2Net leverages the interconnected nature of financial markets by modeling each asset as a node in a graph structure, where node and edge features represent, respectively, univariate and multivariate spot volatilities and volatilities of volatilities. Furthermore, the attention mechanism embedded in the GAT architecture allows SpotV2Net to optimally weigh the influence of specific subgraphs when producing spot volatility forecasts for a specific node. 

The first novelty of this work is to address the rather unexplored topic of intraday volatility forecasting, which is relevant for its economic implications on intraday risk management, high-frequency trading and early detection of market instability conditions. Specifically, it does so by focusing on the actual spot co-volatility matrix, estimated in a non-parametric setting by means of Fourier estimators, rather than by considering parametric volatility estimates or non-parametric integrated volatility estimates computed on short time horizons. 

The second novelty of this work relates to the use of the spot co-volatilities of volatilities as edge features in the graph structure to model the dynamics of the dependence between two nodes and thereby capture spillover effects.  Our findings related to single-step and multi-step forecasts of the spot volatility of the DIJA components suggest that including edge features in the attention mechanism that weighs the importance of subgraphs for predicting the future spot volatility of a specific node yields statistically significant gains in forecasting accuracy, compared to the use of a HAR-type model and other ML models.

Further, the third novelty of our work is to address a typical problem related to ML techniques, that is, interpretability. In this regard,  we use GNNExplainer, a model-agnostic tool designed for graph neural networks, to interpret the forecast provided by the SpotV2Net model. As a result, we obtain insights into the subgraph structures influencing each node's volatility predictions and discover a significant impact of the most recent additions to the DJIA index.  

Note that this paper focuses on predicting individual volatilities, rather than the entire co-volatility matrix, to avoid the curse of dimensionality and reduce computational costs. However, we remark that SpotV2Net's architecture is inherently flexible and thus could be adapted to forecast asset co-volatilities. We leave this extension, which would be particularly beneficial for portfolio optimization and risk management purposes, to future research. 

Another potentially interesting topic for future research may be the development of recurrent networks with GAT layers, which represent an unexplored area of study, to the best of our knowledge. Indeed, integrating a recurrent neural network with the attention mechanism would allow directly capturing the evolution of dynamic graphs over time, rather than relying on static graphs constructed using lagged features. 

\section*{Acknowledgements}
The authors would like to thank the Associate Editor and two anonymous Referees for their valuable comments and suggestions. Furthermore, the authors gratefully acknowledge financial support from the Institut Louis Bachelier 2022 grant `Risk management in times of unprecedented geo-political volatility: a machine learning approach'.

 \bibliographystyle{apalike}
\bibliography{biblio}

\begin{thebibliography}{}

\bibitem[Allaj and Sanfelici, 2023]{ALLAJ20231777}
Allaj, E. and Sanfelici, S. (2023).
\newblock Early warning systems for identifying financial instability.
\newblock {\em International Journal of Forecasting}, 39(4):1777--1803.

\bibitem[Andersen et~al., 2006]{andersen2006volatility}
Andersen, T.~G., Bollerslev, T., Christoffersen, P.~F., and Diebold, F.~X.
  (2006).
\newblock Volatility and correlation forecasting.
\newblock {\em Handbook of Economic Forecasting}, 1:777--878.

\bibitem[Andersen et~al., 2001]{andersen2001distribution}
Andersen, T.~G., Bollerslev, T., Diebold, F.~X., and Ebens, H. (2001).
\newblock The distribution of realized stock return volatility.
\newblock {\em Journal of Financial Economics}, 61(1):43--76.

\bibitem[Bandi et~al., 2008]{bandi2008realized}
Bandi, F.~M., Russell, J.~R., and Yang, C. (2008).
\newblock Realized volatility forecasting and option pricing.
\newblock {\em Journal of Econometrics}, 147(1):34--46.

\bibitem[Barucci and Mancino, 2010]{manbar2010}
Barucci, E. and Mancino, M.~E. (2010).
\newblock Computation of volatility in stochastic volatility models with
  high-frequency data.
\newblock {\em International Journal of Theoretical and Applied Finance},
  13(05):767--787.

\bibitem[Bauwens et~al., 2006]{bauwens2006multivariate}
Bauwens, L., Laurent, S., and Rombouts, J.~V. (2006).
\newblock Multivariate garch models: A survey.
\newblock {\em Journal of Applied Econometrics}, 21(1):79--109.

\bibitem[Becker et~al., 2015]{becker2015selecting}
Becker, R., Clements, A.~E., Doolan, M.~B., and Hurn, A.~S. (2015).
\newblock Selecting volatility forecasting models for portfolio allocation
  purposes.
\newblock {\em International Journal of Forecasting}, 31(3):849--861.

\bibitem[Behrendt and Schmidt, 2018]{BEHRENDT2018355}
Behrendt, S. and Schmidt, A. (2018).
\newblock The twitter myth revisited: Intraday investor sentiment, twitter
  activity and individual-level stock return volatility.
\newblock {\em Journal of Banking \& Finance}, 96:355--367.

\bibitem[Bergstra et~al., 2011]{bergstra2011algorithms}
Bergstra, J., Bardenet, R., Bengio, Y., and K{\'e}gl, B. (2011).
\newblock Algorithms for hyper-parameter optimization.
\newblock {\em Advances in Neural Information Processing Systems}, 24.

\bibitem[Bergstra and Bengio, 2012]{bergstra2012random}
Bergstra, J. and Bengio, Y. (2012).
\newblock Random search for hyper-parameter optimization.
\newblock {\em Journal of Machine Learning Research}, 13(2).

\bibitem[Bollerslev et~al., 1988]{bollerslev1988capital}
Bollerslev, T., Engle, R.~F., and Wooldridge, J.~M. (1988).
\newblock A capital asset pricing model with time-varying covariances.
\newblock {\em Journal of Political Economy}, 96(1):116--131.

\bibitem[Bollerslev et~al., 2019]{bollerslev2019high}
Bollerslev, T., Meddahi, N., and Nyawa, S. (2019).
\newblock High-dimensional multivariate realized volatility estimation.
\newblock {\em Journal of Econometrics}, 212(1):116--136.

\bibitem[Bollerslev et~al., 2009]{BTZ}
Bollerslev, T., Tauchen, G., and Zhou, H. (2009).
\newblock Expected stock returns and variance risk premia.
\newblock {\em The Review of Financial Studies}, 22(11):4463--4492.

\bibitem[Brailsford and Faff, 1996]{brailsford1996evaluation}
Brailsford, T.~J. and Faff, R.~W. (1996).
\newblock An evaluation of volatility forecasting techniques.
\newblock {\em Journal of Banking \& Finance}, 20(3):419--438.

\bibitem[Breiman et~al., 1984]{breiman1984cart}
Breiman, L., Friedman, J., Olshen, R., and Stone, C. (1984).
\newblock Cart.
\newblock {\em Classification and Regression Trees}.

\bibitem[Bucci, 2020]{bucci2020realized}
Bucci, A. (2020).
\newblock Realized volatility forecasting with neural networks.
\newblock {\em Journal of Financial Econometrics}, 18(3):502--531.

\bibitem[Caldeira et~al., 2017]{caldeira2017combining}
Caldeira, J.~F., Moura, G.~V., Nogales, F.~J., and Santos, A.~A. (2017).
\newblock Combining multivariate volatility forecasts: an economic-based
  approach.
\newblock {\em Journal of Financial Econometrics}, 15(2):247--285.

\bibitem[Callot et~al., 2017]{callot2017modeling}
Callot, L.~A., Kock, A.~B., and Medeiros, M.~C. (2017).
\newblock Modeling and forecasting large realized covariance matrices and
  portfolio choice.
\newblock {\em Journal of Applied Econometrics}, 32(1):140--158.

\bibitem[Campisi et~al., 2023]{CAMPISI2023volvol}
Campisi, G., Muzzioli, S., and Baets, B.~D. (2023).
\newblock A comparison of machine learning methods for predicting the direction
  of the us stock market on the basis of volatility indices.
\newblock {\em International Journal of Forecasting}.

\bibitem[Catania and Proietti, 2020]{CATANIA2020volvol}
Catania, L. and Proietti, T. (2020).
\newblock Forecasting volatility with time-varying leverage and volatility of
  volatility effects.
\newblock {\em International Journal of Forecasting}, 36(4):1301--1317.

\bibitem[Chen et~al., 2023]{chen2023deep}
Chen, L., Pelger, M., and Zhu, J. (2023).
\newblock Deep learning in asset pricing.
\newblock {\em Management Science}.

\bibitem[Chen and Robert, 2022]{chen2022multivariate}
Chen, Q. and Robert, C.-Y. (2022).
\newblock Multivariate realized volatility forecasting with graph neural
  network.
\newblock In {\em Proceedings of the Third ACM International Conference on AI
  in Finance}, pages 156--164.

\bibitem[Chen and Guestrin, 2016]{chen2016xgboost}
Chen, T. and Guestrin, C. (2016).
\newblock Xgboost: A scalable tree boosting system.
\newblock In {\em Proceedings of the 22nd ACM SIGKDD International Conference
  on Knowledge Discovery and Data Mining}, pages 785--794.

\bibitem[Chen et~al., 2021]{chen2021asset}
Chen, T.-F., Chordia, T., Chung, S.-L., and Lin, J.-C. (2021).
\newblock Volatility-of-volatility risk in asset pricing.
\newblock {\em The Review of Asset Pricing Studies}, 12(1):289--335.

\bibitem[Cheng et~al., 2022]{cheng2022financial}
Cheng, D., Yang, F., Xiang, S., and Liu, J. (2022).
\newblock Financial time series forecasting with multi-modality graph neural
  network.
\newblock {\em Pattern Recognition}, 121:108218.

\bibitem[Cheng and Li, 2021]{cheng2021modeling}
Cheng, R. and Li, Q. (2021).
\newblock Modeling the momentum spillover effect for stock prediction via
  attribute-driven graph attention networks.
\newblock In {\em Proceedings of the AAAI Conference on Artificial
  Intelligence}, volume~35, pages 55--62.

\bibitem[Christensen et~al., 2023]{christensen2023machine}
Christensen, K., Siggaard, M., and Veliyev, B. (2023).
\newblock A machine learning approach to volatility forecasting.
\newblock {\em Journal of Financial Econometrics}, 21(5):1680--1727.

\bibitem[Christoffersen and Diebold, 2000]{christoffersen2000relevant}
Christoffersen, P.~F. and Diebold, F.~X. (2000).
\newblock How relevant is volatility forecasting for financial risk management?
\newblock {\em Review of Economics and Statistics}, 82(1):12--22.

\bibitem[Corsi, 2009]{corsi2009simple}
Corsi, F. (2009).
\newblock A simple approximate long-memory model of realized volatility.
\newblock {\em Journal of Financial Econometrics}, 7(2):174--196.

\bibitem[Corsi et~al., 2008]{cmpp}
Corsi, F., Mittnik, S., Pigorsch, C., and Pigorsch, U. (2008).
\newblock The volatility of realized volatility.
\newblock {\em Econometric Reviews}, 27(1-3):46--78.

\bibitem[Defferrard et~al., 2016]{defferrard2016convolutional}
Defferrard, M., Bresson, X., and Vandergheynst, P. (2016).
\newblock Convolutional neural networks on graphs with fast localized spectral
  filtering.
\newblock {\em Advances in Neural Information Processing Systems}, 29.

\bibitem[Dempster et~al., 1977]{dempster1977maximum}
Dempster, A.~P., Laird, N.~M., and Rubin, D.~B. (1977).
\newblock Maximum likelihood from incomplete data via the em algorithm.
\newblock {\em Journal of the Royal Statistical Society: Series B
  (methodological)}, 39(1):1--22.

\bibitem[Diao et~al., 2019]{diao2019dynamic}
Diao, Z., Wang, X., Zhang, D., Liu, Y., Xie, K., and He, S. (2019).
\newblock Dynamic spatial-temporal graph convolutional neural networks for
  traffic forecasting.
\newblock In {\em Proceedings of the AAAI conference on Artificial
  Intelligence}, volume~33, pages 890--897.

\bibitem[Diebold and Mariano, 2002]{diebold2002comparing}
Diebold, F.~X. and Mariano, R.~S. (2002).
\newblock Comparing predictive accuracy.
\newblock {\em Journal of Business \& Economic Statistics}, 20(1):134--144.

\bibitem[Ding, 2023]{DING2023joe}
Ding, Y.~D. (2023).
\newblock A simple joint model for returns, volatility and volatility of
  volatility.
\newblock {\em Journal of Econometrics}, 232(2):521--543.

\bibitem[Djanga et~al., 2023]{djanga2023cryptocurrency}
Djanga, E., Cucuringu, M., and Zhang, C. (2023).
\newblock Cryptocurrency volatility forecasting using commonality in intraday
  volatility.
\newblock In {\em Proceedings of the Fourth ACM International Conference on AI
  in Finance}, pages 436--444.

\bibitem[Engle and Kroner, 1995]{engle1995multivariate}
Engle, R.~F. and Kroner, K.~F. (1995).
\newblock Multivariate simultaneous generalized arch.
\newblock {\em Econometric Theory}, 11(1):122--150.

\bibitem[Engle and Sokalska, 2012]{engle2012forecasting}
Engle, R.~F. and Sokalska, M.~E. (2012).
\newblock Forecasting intraday volatility in the us equity market.
  multiplicative component garch.
\newblock {\em Journal of Financial Econometrics}, 10(1):54--83.

\bibitem[Fan et~al., 2019]{fan2019graph}
Fan, W., Ma, Y., Li, Q., He, Y., Zhao, E., Tang, J., and Yin, D. (2019).
\newblock Graph neural networks for social recommendation.
\newblock In {\em The World Wide Web Conference}, pages 417--426.

\bibitem[Fassas and Siriopoulos, 2019]{FASSAS2019333}
Fassas, A.~P. and Siriopoulos, C. (2019).
\newblock Intraday price discovery and volatility spillovers in an emerging
  market.
\newblock {\em International Review of Economics and Finance}, 59:333--346.

\bibitem[Friedman, 2001]{friedman2001greedy}
Friedman, J.~H. (2001).
\newblock Greedy function approximation: a gradient boosting machine.
\newblock {\em Annals of Statistics}, pages 1189--1232.

\bibitem[Gers et~al., 2001]{gers2001applying}
Gers, F.~A., Eck, D., and Schmidhuber, J. (2001).
\newblock Applying lstm to time series predictable through time-window
  approaches.
\newblock In {\em International Conference on Artificial Neural Networks},
  pages 669--676. Springer.

\bibitem[Goldstein et~al., 2023]{goldstein2023hftradstrat}
Goldstein, M., Kwan, A., and Philip, R. (2023).
\newblock High-frequency trading strategies.
\newblock {\em Management Science}, 69:4413–4434.

\bibitem[Golosnoy et~al., 2015]{Golos2015intradayspill}
Golosnoy, V., Gribisch, B., and Liesenfeld, R. (2015).
\newblock Intra-daily volatility spillovers in international stock markets.
\newblock {\em Journal of International Money and Finance}, 53:95--114.

\bibitem[Goodfellow et~al., 2016]{goodfellow2016deep}
Goodfellow, I., Bengio, Y., and Courville, A. (2016).
\newblock {\em Deep learning}.
\newblock MIT press.

\bibitem[Gopal and Chang, 2021]{gopal2021discovering}
Gopal, A. and Chang, C. (2021).
\newblock Discovering supply chain links with augmented intelligence.
\newblock {\em arXiv preprint arXiv:2111.01878}.

\bibitem[Gouri{\'e}roux et~al., 2009]{gourieroux2009wishart}
Gouri{\'e}roux, C., Jasiak, J., and Sufana, R. (2009).
\newblock The wishart autoregressive process of multivariate stochastic
  volatility.
\newblock {\em Journal of Econometrics}, 150(2):167--181.

\bibitem[Granger and Joyeux, 1980]{granger1980introduction}
Granger, C.~W. and Joyeux, R. (1980).
\newblock An introduction to long-memory time series models and fractional
  differencing.
\newblock {\em Journal of Time Series Analysis}, 1(1):15--29.

\bibitem[Gu et~al., 2020]{gu2020empirical}
Gu, S., Kelly, B., and Xiu, D. (2020).
\newblock Empirical asset pricing via machine learning.
\newblock {\em The Review of Financial Studies}, 33(5):2223--2273.

\bibitem[Hamilton et~al., 2017]{hamilton2017inductive}
Hamilton, W., Ying, Z., and Leskovec, J. (2017).
\newblock Inductive representation learning on large graphs.
\newblock {\em Advances in Neural Information Processing Systems}, 30.

\bibitem[Hansen et~al., 2011]{hansen2011model}
Hansen, P.~R., Lunde, A., and Nason, J.~M. (2011).
\newblock The model confidence set.
\newblock {\em Econometrica}, 79(2):453--497.

\bibitem[Heaton et~al., 2017]{heaton2017deep}
Heaton, J.~B., Polson, N.~G., and Witte, J.~H. (2017).
\newblock Deep learning for finance: deep portfolios.
\newblock {\em Applied Stochastic Models in Business and Industry},
  33(1):3--12.

\bibitem[Herskovic et~al., 2016]{herskovic2016common}
Herskovic, B., Kelly, B., Lustig, H., and Van~Nieuwerburgh, S. (2016).
\newblock The common factor in idiosyncratic volatility: Quantitative asset
  pricing implications.
\newblock {\em Journal of Financial Economics}, 119(2):249--283.

\bibitem[Herskovic et~al., 2020]{herskovic2020firm}
Herskovic, B., Kelly, B., Lustig, H., and Van~Nieuwerburgh, S. (2020).
\newblock Firm volatility in granular networks.
\newblock {\em Journal of Political Economy}, 128(11):4097--4162.

\bibitem[Hochreiter and Schmidhuber, 1997]{hochreiter1997long}
Hochreiter, S. and Schmidhuber, J. (1997).
\newblock Long short-term memory.
\newblock {\em Neural Computation}, 9(8):1735--1780.

\bibitem[Huang et~al., 2019]{Huang_Schlag_Shaliastovich_Thimme_2019}
Huang, D., Schlag, C., Shaliastovich, I., and Thimme, J. (2019).
\newblock Volatility-of-volatility risk.
\newblock {\em Journal of Financial and Quantitative Analysis},
  54(6):2423–2452.

\bibitem[Huang et~al., 2021]{huang2021mixgcf}
Huang, T., Dong, Y., Ding, M., Yang, Z., Feng, W., Wang, X., and Tang, J.
  (2021).
\newblock Mixgcf: An improved training method for graph neural network-based
  recommender systems.
\newblock In {\em Proceedings of the 27th ACM SIGKDD Conference on Knowledge
  Discovery \& Data Mining}, pages 665--674.

\bibitem[Jaegle et~al., 2021]{jaegle2021perceiver}
Jaegle, A., Gimeno, F., Brock, A., Vinyals, O., Zisserman, A., and Carreira, J.
  (2021).
\newblock Perceiver: General perception with iterative attention.
\newblock In {\em International Conference on Machine Learning}, pages
  4651--4664. PMLR.

\bibitem[Jawadi et~al., 2015]{jaw2015bidirectional}
Jawadi, F., Louhichi, W., and Idi~Cheffou, A. (2015).
\newblock Intraday bidirectional volatility spillover across international
  stock markets: does the global financial crisis matter?
\newblock {\em Applied Economics}, 47:3633--3650.

\bibitem[Jiang and Luo, 2022]{jiang2022graph}
Jiang, W. and Luo, J. (2022).
\newblock Graph neural network for traffic forecasting: A survey.
\newblock {\em Expert Systems with Applications}, 207:117921.

\bibitem[Katsiampa et~al., 2019]{KATSIAMPA201935}
Katsiampa, P., Corbet, S., and Lucey, B. (2019).
\newblock High-frequency volatility co-movements in cryptocurrency markets.
\newblock {\em Journal of International Financial Markets, Institutions and
  Money}, 62:35--52.

\bibitem[Kim et~al., 2019]{kim2019hats}
Kim, R., So, C.~H., Jeong, M., Lee, S., Kim, J., and Kang, J. (2019).
\newblock Hats: A hierarchical graph attention network for stock movement
  prediction.
\newblock {\em arXiv preprint arXiv:1908.07999}.

\bibitem[Kipf and Welling, 2016]{kipf2016semi}
Kipf, T.~N. and Welling, M. (2016).
\newblock Semi-supervised classification with graph convolutional networks.
\newblock {\em arXiv preprint arXiv:1609.02907}.

\bibitem[LeCun et~al., 1995]{lecun1995convolutional}
LeCun, Y., Bengio, Y., et~al. (1995).
\newblock Convolutional networks for images, speech, and time series.
\newblock {\em The handbook of Brain Theory and Neural Networks},
  3361(10):1995.

\bibitem[Li, 2022]{LI2022energy}
Li, L. (2022).
\newblock The dynamic interrelations of oil-equity implied volatility indexes
  under low and high volatility-of-volatility risk.
\newblock {\em Energy Economics}, 105:105756.

\bibitem[Li and Zhu, 2021]{li2021spatial}
Li, M. and Zhu, Z. (2021).
\newblock Spatial-temporal fusion graph neural networks for traffic flow
  forecasting.
\newblock In {\em Proceedings of the AAAI Conference on Artificial
  Intelligence}, volume~35, pages 4189--4196.

\bibitem[Ligot et~al., 2021]{Ligot2021intradaysmile}
Ligot, S., Gillet, R., and Veryzhenko, I. (2021).
\newblock Intraday volatility smile: Effects of fragmentation and
  high-frequency trading on price efficiency.
\newblock {\em Journal of International Financial Markets, Institutions and
  Money}, 75:101437.

\bibitem[Lin and Taamouti, 2023]{lin2023portfolio}
Lin, W. and Taamouti, A. (2023).
\newblock Portfolio selection under non-gaussianity and systemic risk: A
  machine learning based forecasting approach.
\newblock {\em International Journal of Forecasting}.

\bibitem[Ling and McAleer, 2003]{ling2003asymptotic}
Ling, S. and McAleer, M. (2003).
\newblock Asymptotic theory for a vector arma-garch model.
\newblock {\em Econometric Theory}, 19(2):280--310.

\bibitem[Liu et~al., 2018]{Liu2018tradhfvol}
Liu, F., Pantelous, A.~A., and von Mettenheim, H.-J. (2018).
\newblock Forecasting and trading high-frequency volatility on large indices.
\newblock {\em Quantitative Finance}, 18:737--748.

\bibitem[Liu et~al., 2023]{liu2023preventing}
Liu, J., Cheng, D., and Jiang, C. (2023).
\newblock Preventing attacks in interbank credit rating with selective-aware
  graph neural network.
\newblock In {\em Proceedings of the Thirty-Second International Joint
  Conference on Artificial Intelligence}, pages 6085--6093.

\bibitem[Liu, 2019]{liu2019novel}
Liu, Y. (2019).
\newblock Novel volatility forecasting using deep learning-long short term
  memory recurrent neural networks.
\newblock {\em Expert Systems with Applications}, 132:99--109.

\bibitem[Ma et~al., 2021]{ma2021portfolio}
Ma, Y., Han, R., and Wang, W. (2021).
\newblock Portfolio optimization with return prediction using deep learning and
  machine learning.
\newblock {\em Expert Systems with Applications}, 165:113973.

\bibitem[Maas et~al., 2013]{maas2013rectifier}
Maas, A.~L., Hannun, A.~Y., Ng, A.~Y., et~al. (2013).
\newblock Rectifier nonlinearities improve neural network acoustic models.
\newblock In {\em Proc. ICML}, volume~30, page~3. Atlanta, GA.

\bibitem[Madhusudan and Samit, 2019]{madsam2019intradayrisk}
Madhusudan, K. and Samit, P. (2019).
\newblock Intraday portfolio risk management using var and cvar: A
  cgarch-evt-copula approach.
\newblock {\em International Journal of Forecasting}, 35:699--709.

\bibitem[Malliavin and Mancino, 2002]{malliavin2002fourier}
Malliavin, P. and Mancino, M.~E. (2002).
\newblock Fourier series method for measurement of multivariate volatilities.
\newblock {\em Finance and Stochastics}, 6(1):49--61.

\bibitem[Malliavin and Mancino, 2009]{malliavin2009fourier}
Malliavin, P. and Mancino, M.~E. (2009).
\newblock A fourier transform method for nonparametric estimation of
  multivariate volatility.
\newblock {\em The Annals of Statistics}, 37(4):1983--2010.

\bibitem[Mancino et~al., 2022]{mancino2022asymptotic}
Mancino, M.~E., Mariotti, T., and Toscano, G. (2022).
\newblock Asymptotic normality for the fourier spot volatility estimator in the
  presence of microstructure noise.
\newblock {\em arXiv preprint arXiv:2209.08967}.

\bibitem[Mancino and Recchioni, 2015]{mancino2015fourier}
Mancino, M.~E. and Recchioni, M.~C. (2015).
\newblock Fourier spot volatility estimator: Asymptotic normality and
  efficiency with liquid and illiquid high-frequency data.
\newblock {\em PloS One}, 10(9):e0139041.

\bibitem[Mariotti et~al., 2023]{mariotti2023fromzero}
Mariotti, T., Lillo, F., and Toscano, G. (2023).
\newblock From zero-intelligence to queue-reactive: limit-order-book modeling
  for high-frequency volatility estimation and optimal execution.
\newblock {\em Quantitative Finance}, 23(3):367--388.

\bibitem[Menghani, 2023]{menghani2023efficient}
Menghani, G. (2023).
\newblock Efficient deep learning: A survey on making deep learning models
  smaller, faster, and better.
\newblock {\em ACM Computing Surveys}, 55(12):1--37.

\bibitem[Min et~al., 2021]{min2021stgsn}
Min, S., Gao, Z., Peng, J., Wang, L., Qin, K., and Fang, B. (2021).
\newblock Stgsn—a spatial-temporal graph neural network framework for
  time-evolving social networks.
\newblock {\em Knowledge-Based Systems}, 214:106746.

\bibitem[Monken et~al., 2021]{monken2021graph}
Monken, A., Haberkorn, F., Gopinath, M., Freeman, L., and Batarseh, F.~A.
  (2021).
\newblock Graph neural networks for modeling causality in international trade.
\newblock In {\em The International FLAIRS Conference Proceedings}, volume~34.

\bibitem[Naeem et~al., 2023]{NAEEM2023106677}
Naeem, M.~A., Karim, S., Yarovaya, L., and Lucey, B.~M. (2023).
\newblock Covid-induced sentiment and the intraday volatility spillovers
  between energy and other etfs.
\newblock {\em Energy Economics}, 122:106677.

\bibitem[Natekin and Knoll, 2013]{natekin2013gradient}
Natekin, A. and Knoll, A. (2013).
\newblock Gradient boosting machines, a tutorial.
\newblock {\em Frontiers in neurorobotics}, 7:21.

\bibitem[Nishimura and Sun, 2018]{Nishi2018spillindex}
Nishimura, Y. and Sun, B. (2018).
\newblock The intraday volatility spillover index approach and an application
  in the brexit vote.
\newblock {\em Journal of International Financial Markets, Institutions and
  Money}, 55:241--253.

\bibitem[Panford-Quainoo et~al., 2020]{panford2020bilateral}
Panford-Quainoo, K., Bose, A.~J., and Defferrard, M. (2020).
\newblock Bilateral trade modelling with graph neural networks.
\newblock In {\em ICLR Workshop on Practical ML for Developing Countries}.

\bibitem[Poon and Granger, 2003]{poon2003forecasting}
Poon, S.-H. and Granger, C. W.~J. (2003).
\newblock Forecasting volatility in financial markets: A review.
\newblock {\em Journal of Economic Literature}, 41(2):478--539.

\bibitem[Reisenhofer et~al., 2022]{reisenhofer2022harnet}
Reisenhofer, R., Bayer, X., and Hautsch, N. (2022).
\newblock Harnet: A convolutional neural network for realized volatility
  forecasting.
\newblock {\em arXiv preprint arXiv:2205.07719}.

\bibitem[Rice et~al., 2020]{Rice2020ForVaR}
Rice, G., Wirjanto, T., and Zhao, Y. (2020).
\newblock Forecasting value at risk with intra-day return curves.
\newblock {\em International Journal of Forecasting}, 36:1023--1038.

\bibitem[Rossi and Fantazzini, 2015]{rossi2015long}
Rossi, E. and Fantazzini, D. (2015).
\newblock Long memory and periodicity in intraday volatility.
\newblock {\em Journal of Financial Econometrics}, 13(4):922--961.

\bibitem[Rumelhart et~al., 1985]{rumelhart1985learning}
Rumelhart, D.~E., Hinton, G.~E., Williams, R.~J., et~al. (1985).
\newblock Learning internal representations by error propagation.

\bibitem[Sanfelici and Toscano, 2024]{SanTos24}
Sanfelici, S. and Toscano, G. (2024).
\newblock The {F}ourier-{M}alliavin volatility ({FMV}ol) {MATLAB} library.
\newblock {\em Mathematics and Computers in Simulation}, 226:338--353.

\bibitem[Satchell and Knight, 2011]{satchell2011forecasting}
Satchell, S. and Knight, J. (2011).
\newblock {\em Forecasting volatility in the financial markets}.
\newblock Elsevier.

\bibitem[Scarselli et~al., 2008]{scarselli2008graph}
Scarselli, F., Gori, M., Tsoi, A.~C., Hagenbuchner, M., and Monfardini, G.
  (2008).
\newblock The graph neural network model.
\newblock {\em IEEE Transactions on Neural Networks}, 20(1):61--80.

\bibitem[Shi et~al., 2023]{shi2023edge}
Shi, B., Dong, B., Xu, Y., Wang, J., Wang, Y., and Zheng, Q. (2023).
\newblock An edge feature aware heterogeneous graph neural network model to
  support tax evasion detection.
\newblock {\em Expert Systems with Applications}, 213:118903.

\bibitem[Shwartz-Ziv and Armon, 2022]{shwartz2022tabular}
Shwartz-Ziv, R. and Armon, A. (2022).
\newblock Tabular data: Deep learning is not all you need.
\newblock {\em Information Fusion}, 81:84--90.

\bibitem[Siami-Namini et~al., 2019]{siami2019performance}
Siami-Namini, S., Tavakoli, N., and Namin, A.~S. (2019).
\newblock The performance of lstm and bilstm in forecasting time series.
\newblock In {\em 2019 IEEE International Conference on Big Data}, pages
  3285--3292. IEEE.

\bibitem[Taylor and Xu, 1997]{taylor1997incremental}
Taylor, S.~J. and Xu, X. (1997).
\newblock The incremental volatility information in one million foreign
  exchange quotations.
\newblock {\em Journal of Empirical Finance}, 4(4):317--340.

\bibitem[Toscano, 2022]{Toscano}
Toscano, G. (2022).
\newblock The price-leverage covariation as a measure of the response of the
  leverage effect to price and volatility changes.
\newblock {\em Applied Stochastic Models in Business and Industry},
  38(3):497--511.

\bibitem[Toscano et~al., 2022]{tos2022jfec}
Toscano, G., Livieri, G., Mancino, M.~E., and Marmi, S. (2022).
\newblock Volatility of volatility estimation: Central limit theorems for the
  fourier transform estimator and empirical study of the daily time series
  stylized facts.
\newblock {\em Journal of Financial Econometrics}.

\bibitem[Vaswani et~al., 2017]{vaswani2017attention}
Vaswani, A., Shazeer, N., Parmar, N., Uszkoreit, J., Jones, L., Gomez, A.~N.,
  Kaiser, {\L}., and Polosukhin, I. (2017).
\newblock Attention is all you need.
\newblock {\em Advances in Neural Information Processing Systems}, 30.

\bibitem[Veli{\v{c}}kovi{\'c} et~al., 2017]{velivckovic2017graph}
Veli{\v{c}}kovi{\'c}, P., Cucurull, G., Casanova, A., Romero, A., Lio, P., and
  Bengio, Y. (2017).
\newblock Graph attention networks.
\newblock {\em arXiv preprint arXiv:1710.10903}.

\bibitem[Wang et~al., 2021]{wang2021review}
Wang, J., Zhang, S., Xiao, Y., and Song, R. (2021).
\newblock A review on graph neural network methods in financial applications.
\newblock {\em arXiv preprint arXiv:2111.15367}.

\bibitem[Wilms et~al., 2021]{wilms2021multivariate}
Wilms, I., Rombouts, J., and Croux, C. (2021).
\newblock Multivariate volatility forecasts for stock market indices.
\newblock {\em International Journal of Forecasting}, 37(2):484--499.

\bibitem[Wu et~al., 2021]{wu2021equity2vec}
Wu, Q., Brinton, C.~G., Zhang, Z., Pizzoferrato, A., Liu, Z., and Cucuringu, M.
  (2021).
\newblock Equity2vec: End-to-end deep learning framework for cross-sectional
  asset pricing.
\newblock In {\em Proceedings of the Second ACM International Conference on AI
  in Finance}, pages 1--9.

\bibitem[Wu et~al., 2020a]{wu2020comprehensive}
Wu, Z., Pan, S., Chen, F., Long, G., Zhang, C., and Philip, S.~Y. (2020a).
\newblock A comprehensive survey on graph neural networks.
\newblock {\em IEEE Transactions on Neural Networks and Learning Systems},
  32(1):4--24.

\bibitem[Wu et~al., 2020b]{wu2020connecting}
Wu, Z., Pan, S., Long, G., Jiang, J., Chang, X., and Zhang, C. (2020b).
\newblock Connecting the dots: Multivariate time series forecasting with graph
  neural networks.
\newblock In {\em Proceedings of the 26th ACM SIGKDD International Conference
  on Knowledge Discovery \& Data Mining}, pages 753--763.

\bibitem[Xia et~al., 2022]{xia2022multi}
Xia, L., Huang, C., Xu, Y., Dai, P., and Bo, L. (2022).
\newblock Multi-behavior graph neural networks for recommender system.
\newblock {\em IEEE Transactions on Neural Networks and Learning Systems}.

\bibitem[Xiong et~al., 2015]{xiong2015deep}
Xiong, R., Nichols, E.~P., and Shen, Y. (2015).
\newblock Deep learning stock volatility with google domestic trends.
\newblock {\em arXiv preprint arXiv:1512.04916}.

\bibitem[Xu et~al., 2018]{xu2018powerful}
Xu, K., Hu, W., Leskovec, J., and Jegelka, S. (2018).
\newblock How powerful are graph neural networks?
\newblock {\em arXiv preprint arXiv:1810.00826}.

\bibitem[Xue and Gençay, 2012]{Xue2012freqclust}
Xue, Y. and Gençay, R. (2012).
\newblock Trading frequency and volatility clustering.
\newblock {\em Journal of Banking and Finance}, 36:760--773.

\bibitem[Ying et~al., 2018]{ying2018graph}
Ying, R., He, R., Chen, K., Eksombatchai, P., Hamilton, W.~L., and Leskovec, J.
  (2018).
\newblock Graph convolutional neural networks for web-scale recommender
  systems.
\newblock In {\em Proceedings of the 24th ACM SIGKDD International Conference
  on Knowledge Discovery \& Data Mining}, pages 974--983.

\bibitem[Ying et~al., 2020]{ying2020time}
Ying, X., Xu, C., Gao, J., Wang, J., and Li, Z. (2020).
\newblock Time-aware graph relational attention network for stock
  recommendation.
\newblock In {\em Proceedings of the 29th ACM International Conference on
  Information \& Knowledge Management}, pages 2281--2284.

\bibitem[Ying et~al., 2019]{ying2019gnnexplainer}
Ying, Z., Bourgeois, D., You, J., Zitnik, M., and Leskovec, J. (2019).
\newblock Gnnexplainer: Generating explanations for graph neural networks.
\newblock {\em Advances in Neural Information Processing Systems}, 32.

\bibitem[Zhang et~al., 2022a]{zhang2022graph}
Zhang, C., Pu, X.~S., Cucuringu, M., and Dong, X. (2022a).
\newblock Graph-based methods for forecasting realized covariances.
\newblock {\em Available at SSRN}.

\bibitem[Zhang et~al., 2023a]{zhang2023graph}
Zhang, C., Pu, X.~S., Cucuringu, M., and Dong, X. (2023a).
\newblock Graph neural networks for forecasting realized volatility with
  nonlinear spillover effects.
\newblock {\em Available at SSRN}.

\bibitem[Zhang et~al., 2023b]{zhang2023volatility}
Zhang, C., Zhang, Y., Cucuringu, M., and Qian, Z. (2023b).
\newblock Volatility forecasting with machine learning and intraday
  commonality.
\newblock {\em Journal of Financial Econometrics}.

\bibitem[Zhang et~al., 2022b]{zhang2022research}
Zhang, W., Chen, Z., Miao, J., and Liu, X. (2022b).
\newblock Research on graph neural network in stock market.
\newblock {\em Procedia Computer Science}, 214:786--792.

\bibitem[Zhang et~al., 2020]{zhang2020deep}
Zhang, Z., Zohren, S., and Roberts, S. (2020).
\newblock Deep learning for portfolio optimization.
\newblock {\em The Journal of Financial Data Science}.

\bibitem[Zhao et~al., 2022]{zhao2022combining}
Zhao, Y., Wei, S., Guo, Y., Yang, Q., Chen, X., Li, Q., Zhuang, F., Liu, J.,
  and Kou, G. (2022).
\newblock Combining intra-risk and contagion risk for enterprise bankruptcy
  prediction using graph neural networks.
\newblock {\em arXiv preprint arXiv:2202.03874}.

\bibitem[Zhou et~al., 2020]{zhou2020graph}
Zhou, J., Cui, G., Hu, S., Zhang, Z., Yang, C., Liu, Z., Wang, L., Li, C., and
  Sun, M. (2020).
\newblock Graph neural networks: A review of methods and applications.
\newblock {\em AI Open}, 1:57--81.

\bibitem[Zhu et~al., 2023]{zhu2023forecasting}
Zhu, H., Bai, L., He, L., and Liu, Z. (2023).
\newblock Forecasting realized volatility with machine learning: Panel data
  perspective.
\newblock {\em Journal of Empirical Finance}, 73:251--271.

\end{thebibliography}

\appendix

\section{Alternative models}\label{sec:appendix_alternativemodels}

\subsection{Heterogeneous Autoregressive (HAR) Spot Model}
We adapt the HAR model by \cite{corsi2009simple} to forecast intraday spot volatilities. The resulting model, which we term HAR-Spot, assumes a linear relationship between the 30-minute-ahead spot volatility and three regressors: the current spot volatility, the average spot volatility between lag $1$ (30 minutes) and lag $7$ (3.5 hours) and the average spot volatility between lag $8$ (4 hours) and lag $13$ (6.5 hours, that is, 1 trading day). This way, in the spirit of the original model by \cite{corsi2009simple}, the HAR-Spot model captures the persistence of the spot volatility in a parsimonious way. Specifically, we implement the following panel version of the HAR-Spot model. For the $i$-th asset, the HAR-Spot model reads as

\begin{equation*}\label{Eq:extendedHAR}
\begin{aligned}
\widehat{V}_{i,b+1} = & \ \mu + \phi_1 \widehat{V}_{i,b} + \phi_2 \left(\frac{1}{7} \sum_{l=1}^7 \widehat{V}_{i,b-l}\right) + \phi_3 \left(\frac{1}{7} \sum_{l=8}^{13} \widehat{V}_{i,b-l}\right) \\
& + \sum_{k \neq i} \left[ \theta_1 \widehat{V}_{k,t} + \theta_2 \left(\frac{1}{7} \sum_{l=1}^7 \widehat{V}_{k,b-l}\right) + \theta_3 \left(\frac{1}{7} \sum_{l=8}^{13} \widehat{V}_{k,b-l}\right) \right] + \epsilon_{i,b},
\end{aligned}
\end{equation*}

where $\sum_{k \neq i} \left[ \cdots \right]$ denotes the contribution from the lagged volatilities of all other assets $k \neq i$.

%%%%%%%%%%%%%%%%%%%%%%%%%%%%%%%%%%%%%%%%%%%%%%%%%%%%%%%%%%%%%%%%%%%%%%%%%%%%%%%%%%%%
\subsection{Extreme Gradient Boosting (XGB)}

Ensembles of weak learners, such as regression trees \citep{breiman1984cart}, can effectively handle non-linear interactions between predictors. Specifically, XGB \citep{chen2016xgboost} is a tree-based ensemble algorithm that has proven effective in addressing prediction problems based on tabular data \citep{shwartz2022tabular}. Hence, our approach is to input the regressors to such an algorithm in the same form we input them to fit the HAR-Spot model. Letting $u_{i,b}$ denote the vector of input features for the $i$-th asset at time $\tau_b$,
 the XGB model is specified as
\begin{equation*}
     F_i(u_{i,b})=\sum_{j=1}^J f_j(u_{i,b}), \quad f_j \in \mathcal{R},
\end{equation*}
  where $J$ is the number of estimators composing the ensemble and $\mathcal{R}$ is the space of regression trees. The tree ensemble model is trained sequentially. The boosting technique \citep{friedman2001greedy} indicates that trees are added to minimize the errors made by previously fitted trees until no further improvements are achieved. The optimization procedure builds trees as a forward mechanism, where every step reduces the error of the previous iteration. We initialize the ensemble with a single regression tree and then iteratively add new trees that minimize the error made by the previous tree by gradient descent.

We chose to train XGB using the same inputs as for the HAR-Spot model because training the latter using all the inputs considered in the node and edge features of SpotV2Net resulted in lower performance. 
This choice aligns with findings in the literature suggesting that when the feature space is too large, gradient boosting techniques, including XGB, are more prone to overfitting \cite{natekin2013gradient}.  Additionally, limiting the number of features to just the volatilities, which represent the objects of forecasting, simplifies the multi-step forecasting process. Like HAR-Spot, XGB performs multi-step forecasts recursively. Hence, including additional features in the input space, such as, e.g., the co-volatilities, would require XGB to recursively forecast not only volatilities but also the additional features.

%%%%%%%%%%%%%%%%%%%%%%%%%%%%%%%%%%%%%%%%%%%%%%%%%%%%%%%%%%%%%%%%%%%%%%%%%%%%%%%%%%%%
\subsection{Long Short-term Memory  (LSTM) Neural Network}
Recurrent Neural Networks (RNNs), see \cite{hochreiter1997long}, are models designed to retain information about sequential data, making them particularly effective in the financial context where one usually forecasts future values based on historical data. A performing version of an RNN is the LSTM network \citep{hochreiter1997long}, which has contributed to significant breakthroughs across several fields, including natural language processing, time series forecasting and generative modeling \citep{siami2019performance,gers2001applying}. 

The LSTM architecture proposes a memory block to capture long-term dependencies more effectively. Within each LSTM unit, a standard transformation of the original input occurs. Consider the vector of input features $\mathbf{x}_{\tau_b}$ at time $\tau_b$, which comprises all companies' univariate volatilities and co-volatilities. The LSTM model accepts these vectors $\mathbf{x}_{\tau_b}$ up to a given lag $L$. Note that we input only spot volatilities and co-volatilities time series in the LSTM algorithm because our experiments suggest that also inputting univariate and multivariate co-volatilities of volatilities and their lags worsens the model's performance. We attribute this performance decrease to the large dimensionality of the input size. In contrast, SpotV2Net uses this information efficiently by distributing it into a graph structure with node and edge features. LSTM has no knowledge of such structure and thus infers relationships by combining information through different layers, which proves to be limited in our empirical experiment.

The transformation performed by the LSTM on the vectors $\mathbf{x}_{\tau_b}$ at time $\tau_b$ reads as
 
\begin{align*}
\mathbf{f}_{\tau_b} & =\sigma_g\left(\mathbf{W}_f \mathbf{x}_{\tau_b}+U_f \mathbf{h}_{\tau_{b-1}}+\mathbf{b}_f\right) \\
\mathbf{i}_{\tau_b} & =\sigma_g\left(\mathbf{W}_i \mathbf{x}_{\tau_b}+U_i \mathbf{h}_{\tau_{b-1}}+\mathbf{b}_i\right) \\
\mathbf{o}_{\tau_b} & =\sigma_g\left(\mathbf{W}_o \mathbf{x}_{\tau_b}+U_o \mathbf{h}_{\tau_{b-1}}+\mathbf{b}_o\right) \\
\tilde{\mathbf{c}}_t & =\sigma_c\left(\mathbf{W}_c \mathbf{x}_t+U_c \mathbf{h}_{\tau_{b-1}}+\mathbf{b}_c\right) \\
\mathbf{c}_{\tau_b} & =\mathbf{f}_{\tau_b} \circ \mathbf{c}_{\tau_{b-1}}+\mathbf{i}_{\tau_b} \circ \tilde{\mathbf{c}}_{\tau_b} \\
\mathbf{h}_{\tau_b} & =\mathbf{o}_{\tau_b} \circ \sigma_h\left(\mathbf{c}_{\tau_b}\right)
\end{align*}
where $\mathbf{x}_{\tau_b}$ is the input vector, $\mathbf{f}_{\tau_b}$ is the forget gate's activation vector, $\mathbf{i}_{\tau_b}$ is the update gate's activation vector, $\mathbf{o}_{\tau_b}$ is the output gate's activation vector, $\tilde{\mathbf{c}}_{\tau_b}$ is the cell input activation vector, $\mathbf{c}_{\tau_b}$ is the cell state vector and $\mathbf{h}_{\tau_b}$ is the hidden state vector, i.e. the output vector of the LSTM unit. Besides, $\circ$ is the Hadamard product function, $\sigma_g$ is the sigmoid function and $\sigma_c$ and $\sigma_h$ are hyperbolic tangent functions. Finally, $\mathbf{W}_{(\cdot)}$ and $\mathbf{b}_{(\cdot)}$ refer to weight matrices and bias vectors that need to be estimated by training the model, while $\mathbf{U}_{(\cdot)}$ refers to the weight matrices that are applied to the hidden state vector from the previous time step $\mathbf{h}_{\tau_{b-1}}$.

%%%%%%%%%%%%%%%%%%%%%%%%%%%%%%%%%%%%%%%%%%%%%%%%%%%%%%%%%%%%%%%%%%%%%%%%%%%%%%%%%%%%

\section{Hyperparameters}\label{Sec:appendix_hyperparameters}

\setlength{\floatsep}{5pt}
\setlength{\textfloatsep}{5pt}
\setlength{\intextsep}{5pt}

For the SpotV2Net, XGB and LSTM models, we utilized the \href{https://pytorch-geometric-temporal.readthedocs.io/en/latest/index.html}{Optuna} framework implemented in Python, which performs a search over a predefined grid of hyperparameters. This technique is commonly employed for optimizing neural network-based models. In our hyperparameter optimization process with Optuna, we employ the Tree-structured Parzen Estimator (TPE) algorithm as our sampler \citep{bergstra2011algorithms,bergstra2012random}. The TPE algorithm enhances the efficiency of the search by using a Bayesian optimization technique. During each trial, for each parameter, TPE fits one Gaussian Mixture Model (GMM) \citep{dempster1977maximum} $l(x)$ to the set of parameter values associated with the best objective values and another GMM $g(x)$ to the remaining parameter values. It then chooses the parameter value $x$ that maximizes the ratio $\frac{l(x)}{g(x)}$. By choosing the hyperparameter values that maximize this ratio, TPE focuses the search on regions where the likelihood of improving the objective function is higher. This approach ensures that our hyperparameter tuning is both effective and computationally efficient.

The grid for Optuna is constructed by specifying a meaningful set of possible values for each hyperparameter, which we report together with the optimized values in Tabs. \ref{tab:GAT_hyperparameters}, \ref{tab:LSTM_hyperparameters} and \ref{tab:XGB_hyperparameters} for the SpotV2Net, LSTM and XGB models, respectively. Due to memory constraints on the NVIDIA GeForce RTX 2080 Ti with 12 GB of memory that we used to train the deep learning models, we were unable to expand the grid search to include configurations with more than three layers, larger layer sizes, higher number of heads, or larger batch updates, as these modifications would exceed the available computational resources. The optimal values reported are the result of this search process, indicating the configuration that achieved the best performance on the validation set, as shown in Tab. \ref{tab:data_splits}. The hyperparameters are optimized for all models by evaluating their performance on a validation set, ensuring that the test set remains unseen during the training and tuning phases. This strategy prevents any bias towards the test set and ensures that our evaluation metrics reflect the models' ability to generalize to new data.

\vspace{0.25cm}

\begin{table}[h!]
\centering
\begin{tabular}{p{5cm} p{4cm} p{2cm} p{2cm}}
\toprule
\textbf{Hyperparameter} & \textbf{Grid search options} & \textbf{Single-step opt. val.} & \textbf{Multi-step opt. val.} \\
\midrule
Loss Function & - & MSE & MSE\\[0.4ex]
Activation Function ($\sigma$) & ReLu, Tanh, Sigmoid & ReLu & ReLu \\[0.4ex]
Batch Size & 32, 64, 128, 256 & 128 & 128 \\[0.4ex]
Concatenate Heads & True, False & True & True \\[0.4ex]
Number Hidden Layers & 1, 2, 3 & 2 & 2 \\[0.4ex]
Dimension Hidden Layers ($M^\prime$) & from 50 to 600 (step 50) & [400, 200] & [400, 400] \\[0.4ex]
Dropout (Architecture) & 0.0 to 0.7 (step 0.1) & 0.1 & 0.2 \\[0.4ex]
Dropout (Attention) & 0.0 to 0.7 (step 0.1) & 0.1 & 0.0 \\[0.4ex]
Learning Rate & $\{1, 2.5, 5\} \times 10^{-2 \text{ to } -5}$ & $1 \times 10^{-4}$  & $5 \times 10^{-5}$ \\[0.4ex]
Negative Slope LeakyReLu ($c$) & 0.05, 0.1, 0.2, 0.5, 0.6, 0.8 & 0.1 & 0.1 \\[0.4ex]
Epochs & 60 to 200 (step 20) & 120 & 120 \\[0.4ex]
Number of Heads ($K$) & 2, 3, 4, 5, 6, 7 & 4 & 5 \\[0.4ex]
Optimizer & RMSProp, Adam, AdamW & AdamW & AdamW \\[0.4ex]
Output Dim & - & 1 & 14 \\[0.4ex]
Number of Lags ($L$) & 14 to 112 (step 14) & 42 & 42 \\[0.4ex]
\bottomrule
\end{tabular}
\caption{Hyperparameter grid search options and optimized values for the single-step and multi-step SpotV2Net models.}
\label{tab:GAT_hyperparameters}
\end{table}

\begin{table}[h!]
\centering
\begin{tabular}{p{4.5cm} p{6cm} p{2cm}}
\toprule
\textbf{Hyperparameter} & \textbf{Grid search options} & \textbf{Opt. val.} \\
\midrule
Subsample & 0.6, 0.7, 0.8, 0.9, 1 & 0.7 \\
Regularization $\lambda$ & 0.5, 1, 1.5, 2 & 1.5 \\
Regularization $\alpha$ & 0, 0.1, 0.5, 1 & 0 \\
Number of Estimators ($J$) & 50, 100, 150, 200, 300, 400, 500, 600 & 400 \\
Minimum Child Weight & 1, 3, 5, 7 & 5 \\
Max Depth & 3, 4, 5, 6, 7, 8 & 5 \\
Learning Rate & 0.001, 0.01, 0.05, 0.1, 0.2 & 0.2 \\
Gamma & 0, 0.05, 0.1, 0.2, 0.3, 0.4 & 0 \\
Colsample by Tree & 0.5, 0.6, 0.7, 0.8, 0.9, 1 & 1 \\
\bottomrule
\end{tabular}
\caption{Hyperparameter grid search options and optimized values for the XGB model.}
\label{tab:XGB_hyperparameters}
\end{table}

\begin{table}[h]
\centering
\begin{tabular}{p{5cm} p{4cm} p{2cm} p{2cm}}
\toprule
\textbf{Hyperparameter} & \textbf{Grid search options} & \textbf{Single-step opt. val.} & \textbf{Multi-step opt. val.} \\
\midrule
Loss Function & - & MSE & MSE\\[0.4ex]
Batch Size & 32, 64, 128, 256 &  64 & 64 \\[0.4ex]
Dropout (Architecture) & 0.0 to 0.7 (step 0.1)  & 0.4 & 0.3 \\[0.4ex]
Number Hidden Layers & 1, 2, 3 & 2 & 2 \\[0.4ex]
Dimension Hidden Layers & from 50 to 600 (step 50) & [400, 200] & [400, 400] \\[0.4ex]
Learning Rate & $\{1, 2.5, 5\} \times 10^{-2 \text{ to } -5}$ & $5 \times 10^{-4}$  & $5 \times 10^{-4}$ \\[0.4ex]
Optimizer & RMSProp, Adam, AdamW & AdamW & AdamW \\[0.4ex]
Output Dim & - & 1 & 14 \\[0.4ex]
\bottomrule
\end{tabular}
\caption{Hyperparameter grid search options and optimized values for the single-step and multi-step LSTM models.}
\label{tab:LSTM_hyperparameters}
\end{table}

\end{document}